\documentclass[10pt,twocolumn,superscriptaddress,aps,pra,balancelastpage]{revtex4-2}
\usepackage[utf8]{inputenc}
\usepackage{amsmath,amsfonts,amssymb,fullpage,color,graphicx,xcolor,physics}
\usepackage[colorlinks]{hyperref}
\usepackage{CJKutf8}

\begin{document}
\begin{CJK}{UTF8}{gbsn}
\title{Uncovering measurement-induced entanglement via directional adaptive dynamics and incomplete information}

\author{Yu-Xin Wang (王语馨)}
\email{yxwang@uchicago.edu}
\affiliation{Pritzker School of Molecular Engineering, University of Chicago, Chicago, IL 60637}
\affiliation{Joint Center for Quantum Information and Computer Science, University of Maryland, College Park, MD 20742, USA}
\author{Alireza Seif}
\affiliation{Pritzker School of Molecular Engineering, University of Chicago, Chicago, IL 60637}
\author{Aashish A. Clerk}
\affiliation{Pritzker School of Molecular Engineering, University of Chicago, Chicago, IL 60637}
\begin{abstract} 
The rich entanglement dynamics and transitions exhibited by monitored quantum systems typically only exist in the conditional state, making observation extremely difficult.  In this work we construct a general recipe for mimicking the conditional entanglement dynamics of a monitored system in a corresponding measurement-free dissipative system involving directional interactions between the original system and a set of auxiliary register modes.  This mirror setup autonomously implements a measurement-feedforward dynamics that effectively retains a coarse-grained measurement record.  We illustrate our ideas in a bosonic system featuring a competition between entangling measurements and local unitary dynamics, and also discuss extensions to qubit systems and truly many-body systems.    
\end{abstract}
\maketitle
\end{CJK}

\begin{figure}[t]
    \centering
    \includegraphics[width=\columnwidth]{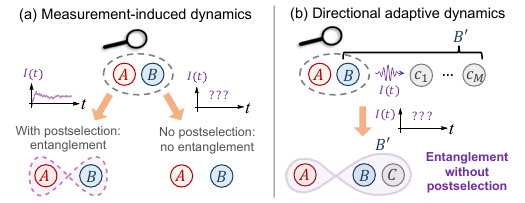}
    \caption{Adaptive approach for mimicking entanglement dynamics of a monitored system.  (a) A collective measurement creates conditional entanglement between $A$ and $B$, whereas the unconditional state (averaged over measurement outcomes) is unentangled.  (b)  
    By applying a conditional feedforward operation on auxiliary register modes $c _{j}$ ($j=1,2,\ldots,M$), partial information from the measurement record is retained.  This can be enough to maintain entanglement in the unconditional state (now between $A$ and the composite subsystem of ($B$, $c _{j}$)). This dynamics can be realized fully autonomously (without any measurement) using an engineered dissipative process 
    (c.f.~Eq.~\eqref{eq:qme.meas.ff.nmodes}).
    }
    \label{fig:schematic}
\end{figure}

{\it Introduction-- }
Entanglement is a unique feature of quantum systems, and studying its dynamics in complex systems has both fundamental and practical motivations.  
To wit, there is immense interest in understanding different phases of entanglement generation in systems having a competition between Hamiltonian and measurement-induced dynamics (see e.g.~\cite{Nahum2019,Fisher2019,Pixley2020,BaoYM2020,Altman2020,Barkeshli2021,Schiro2021,Hafezi2021,Diehl2021,Muller2021,Barkeshli2021PRL,MunozArias2023,Buchhold2022,Ashida2024,Vasseur2022,Vijay2023}). A common feature here is that entanglement generation is contingent on knowing the measurement results, i.e.,~it only exists at the level of individual measurement trajectories (see Fig.~\ref{fig:schematic}(a)). Conversely, the average state (averaged over all measurement outcomes) is typically highly mixed and unentangled. As such, direct detection of novel entanglement dynamics and transitions would seem to require postselection over measurement records, posing formidable challenges for scalable experimental implementation~\cite{Minnich2023}.

Various ideas have been suggested to tackle this postselection problem~\cite{Khemani2021,Ojanen2023,Gullans2023,Altman2023,Fazio2023,Wilson2022,Diehl2022,Turkeshi2023PRL_AbsorbingState,Khemani2022,Turkeshi2023,Huse2020,Fisher2022}, with some corresponding experimental implementations~\cite{Monroe2022,Google2023MIE}.
Many of these approaches focus on measuring a proxy quantity (i.e.~not system entanglement directly), or study a transition in the efficiency of using feedback-assisted dynamics to stabilize a pre-selected target state (with this transition serving as a proxy for the actual
measurement-induced entanglement phase transitions (MIPT) of interest~\cite{Wilson2022,Diehl2022,Turkeshi2023PRL_AbsorbingState,Khemani2022}). While those approaches do not require postselection, one might worry that the transitions in feedback-assisted dynamics might be distinct and only loosely related to the original entanglement phase transition~\cite{Nandkishore2022,Nahum2023,Turkeshi2023PRL_AbsorbingState,Khemani2022,ChenXiao2023,Turkeshi2023,Wilson2023}.

Given this, it would be highly desirable to have a fully deterministic, postselection free protocol whose entanglement dynamics (and not some proxy quantity) quantitatively reproduces transitions in the trajectory-level entanglement of a 
measured system.  
In this work, we introduce such a strategy.   
We start with a system of interest, where a combination of Hamiltonian dynamics and continuous measurements leads to interesting features in the \textit{postselected} entanglement dynamics (Fig.~\ref{fig:schematic}(a)).  To access this physics, we propose studying
a distinct, modified setup that includes the original system (the ``target") and its Hamiltonian dynamics.  We now have no explicit measurements, but instead introduce one or more ``register" systems, and implement {\it dissipative} dynamics that {\it directionally} couples the target to these registers.  This dissipative dynamics is constructed to autonomously mimic an adaptive process involving continuous measurement on the target, followed by conditional feedforward driving of the register \cite{Wiseman1994,Wiseman2009book,Metelmann2017} (Fig.~\ref{fig:schematic}(b)).

Unlike protocols based on deferred measurement~\cite{Monroe2022,Nandkishore2022}, our setup does not generate any entanglement between the register and the system.  Further, it only retains a vanishingly small fraction of the information that would be contained in an actual measurement record.   
Despite these caveats, we find that if one considers an appropriate bipartition (where part of the target is grouped with the registers), 
the entanglement of the constructed dissipative dynamics  
can quantitatively capture features of entanglement generation in the original monitored, postselected system.
Our approach thus has the potential of providing a new and powerful method for accessing MIPTs.

While our scheme is general, we focus here on a relatively unexplored setting, where there is a competition between collective entanglement-generating measurements and a set of local Hamiltonians that can either enhance or suppress entanglement creation.
This is the opposite of what is typically studied, where the focus is instead on the interplay between entangling unitaries and competing measurements~\cite{Nahum2019,Fisher2019,Pixley2020,BaoYM2020,Altman2020,Barkeshli2021,Schiro2021,Hafezi2021,Diehl2021,Muller2021,Barkeshli2021PRL,MunozArias2023,Buchhold2022}.
Our adaptive-dynamics setup generates entanglement growth that exhibits almost the same parametric dependence as the original measured, postselected system.  As an application, we show that for two bosonic modes, the adaptive dynamics generates unlimited entanglement with logarithmic growth in time. We also show that our measurement-free approach to realizing MIPTs is directly applicable to many-body settings, as well as nonlinear systems such as qubits and qudits. 
We end by discussing experimental implementations, which are within reach using current physical platforms.

{\it Basic setup-- }While our scheme is suitable for many-body systems, to illustrate the basic ideas we start by analyzing the simplest non-trivial example:  a target system consisting of two bosonic modes, $ a $ and $ b $.  We define quadrature operators in terms of annihilation operators $ \hat a $ and $\hat b $ as 
$\hat {x} _{ m } \equiv  
(\hat { m } + 
\hat { m } ^\dag  ) 
/ \sqrt{2} $, ($ m =a,b $).    
Consider the postselected dynamics due to competing processes of a continuous entangling measurement of the collective quadrature $\hat {x} _{+} \equiv 
\left (\hat {x} _{a} + 
\hat {x} _{b} \right) / \sqrt{2} $, and a local Hamiltonian
\begin{align}
\label{eq:h.qmfs.perturb}
&  \hat H _{\mathrm{det}}
= (\omega + \delta \omega ) \hat a ^{\dag}
\hat a 
+ ( - \omega + \delta \omega ) \hat b ^{\dag}
\hat b
. 
\end{align}
It is well known that large many-body lattice systems can exhibit distinct phases of entanglement dynamics, characterized by entanglement either growing linearly or logarithmically in time (volume law or critical phases), or saturating (area law phases).  Similar regimes can also exist in few mode bosonic systems, something that is enabled by their infinite-dimensional Hilbert space. 
We find that depending on parameters, the measurement-induced, inter-mode entanglement generation in 
the two-mode system above can exhibit drastically different asymptotic regimes, due to a competition between the entangling measurement and non-entangling Hamiltonian dynamics.

\begin{figure}[t]
    \centering
    \includegraphics[width=\columnwidth]{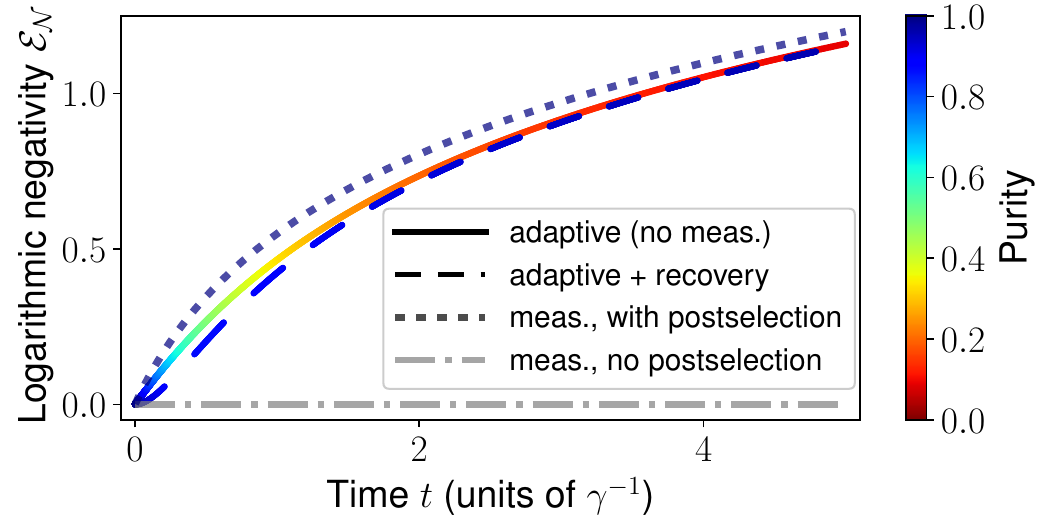}
    \caption{Preservation of postselected entanglement generation via adaptive dynamics. Dotted curve: Conditioned entanglement growth between modes $ a $ and $ b $, quantified by logarithmic negativity $\mathcal{E} _{\mathcal{N}} 
^{\mathrm{(ps)}} $, of the postselected state generated by continuous measurement of the quadrature $\hat {x} _{+} \equiv 
\left (\hat {x} _{a} + 
\hat {x} _{b} \right) / \sqrt{2} $ (see Eq.~\eqref{eq:meas.cond.xp.ent}). 
    Solid and dashed curves: Deterministic entanglement between modes 
    $a $ and $(b, c) $ as generated by the postselection-free dissipative dynamics in Eq.~\eqref{eq:qme.meas.ff}, for cases with (solid curve) and without (dashed curve) recovery via projection measurement of mode $c$ quadrature 
    $\hat \pi $ and conditional feedback. Coloring of the curves (except gray) corresponds to state purity. Parameter: $\eta =1$.  
    }
    \label{fig:mff_ent}
\end{figure}

To quantify entanglement, we will use the logarithmic negativity $ \mathcal{E} ^{\mathrm{(ps)}} _{\mathcal{N}} $ of the postselected state. In the absence of a Hamiltonian 
($\omega = \delta \omega =0$) and for an initial vacuum state, we obtain unbounded entanglement growth (see Fig.~\ref{fig:mff_ent})
\begin{align}
\label{eq:meas.cond.xp.ent}
& \mathcal{E} _{\mathcal{N}} 
^{\mathrm{(ps)}}
=  (1/2)
\log  (1+ 2
\gamma t)
    , 
\end{align}
where $\gamma$ and $t$ denote measurement rate and evolution time, respectively~\footnote{In this specific case, as the dynamics is Gaussian, every trajectory has the same entanglement.} (see~\cite{SI} for details). The logarithmic temporal dependence of entanglement generation is known to be a feature of conformal invariance~\cite{Calabrese2004}, and emerges in nonequilibrium systems including many-body localized phases~\cite{Prelovsek2008,Moore2012,Serbyn2019} and measurement-induced dynamics~\cite{Ludwig2020,Fisher2021}. Here, the logarithmic scaling arises due to quantum-nondemolition (QND) measurement-induced squeezing (see~\cite{Rigol2018} for a Hamiltonian version of this effect). 

If we now include local Hamiltonian dynamics generated by Eq.~\eqref{eq:h.qmfs.perturb}, we find that entanglement growth is hindered whenever $\delta \omega \ne 0$: there is no longer any unbounded growth.  In contrast (and somewhat surprisingly), the local Hamiltonian dynamics can {\it enhance} entanglement generation when $\delta \omega = 0$ and $\omega$ is non-zero.    
We stress these different regimes of entanglement dynamics can only be seen with postselection:  if one instead averages over all the measurement outcomes, the resulting unconditioned state is unentangled at all times (irrespective of parameters).

We now seek to reproduce the conditional entanglement dynamics of this measured system in  a second postselection-free (and potentially measurement-free) setup.  
Making use of the general recipe laid out in the introduction, we consider an adaptive process where a coarse grained version of the measurement record obtained from monitoring $\hat x _{+}$ is retained in a set of auxiliary register bosonic modes $c_j$ ($j =1, 2,\ldots, M$).   This could be achieved by partitioning the total evolution time $t_f$ into $M$ equal intervals, and in each interval, using the measurement record to linearly force one of the $M$ register modes (see Fig.~\ref{fig:schematic}(b)).  As shown in~\cite{Wiseman1994,Wiseman2009book,Metelmann2017}, the \textit{unconditioned} evolution from such a process in the zero-delay limit is described by the master equation 
\begin{align}
\label{eq:qme.meas.ff.nmodes}
&  (d \hat \rho  / dt) = 
- i [ \hat H _{\mathrm{det}}
, \hat \rho ] +
\gamma \sum _{j =1} ^{ M } 
f_{j} (t; t_{f}) \mathcal{L} 
_{ \hat x _{+}  
\to \eta \hat y _{j} }  
\hat \rho 
, 
\end{align}
where $f_{j} (t; t_{f}) 
\equiv 
\Theta (t-\frac{j-1}{M} t_{f}) 
\Theta (\frac{j}{M} t_{f} - t ) $ 
ensures the measurement record from the $j$th interval drives the register mode $\hat{c}_j$ ($\Theta (\cdot)$ is the Heaviside step function). We define the Lindbladians $ \mathcal{L} 
_{ \hat x _{+}  
\to \eta \hat y _{j} } $ as 
\begin{align}
\label{eq:qme.meas.ff}
&  \mathcal{L} 
_{ \hat x _{+}  
\to \eta \hat y _{j} } 
\hat \rho  
\equiv 
- i 
[\eta \hat x _{+} 
\hat y _{j} , 
\hat{ \rho} ] 
+  
\mathcal{D}\left[ 
\hat x _{+} 
- i \eta \hat y _{j} \right ] 
\hat{ \rho}  
, 
\end{align}
with $\hat y _{j} \equiv  (\hat {c} _{j} + 
\hat {c} _{j} ^{\dag}  ) / \sqrt{2} $ a quadrature of register $j$, $\eta $ a constant quantifying feedforward strength, and $  \mathcal{D} [\hat{O}  ]  
\hat{\rho} \equiv  
 (\hat{O}  \hat{\rho} \hat{O} ^\dagger 
- \{\hat{O} ^\dagger \hat{O}
, \hat{\rho}\} /2  ) $ denoting the standard Lindblad dissipator~\cite{Lindblad1976,Sudarshan1976}. 

While we have motivated Eq.~\eqref{eq:qme.meas.ff} by considering unconditional evolution in a setup with explicit measurements and feedforward, identical dynamics could be achieved
without any measurements at all:  one could engineer an autonomous dissipative process that yields the same master equation (using the tools of reservoir engineering, see e.g.~\cite{Metelmann2015}).  This then provides a potential measurement-free route for probing the entanglement physics of the original monitored system.  We stress that the system-only dynamics is completely unaffected by the feedforward step, i.e., if we trace out the auxiliary registers in Eq.~\eqref{eq:qme.meas.ff.nmodes}, the system dynamics faithfully recovers the unconditioned evolution associated with measuring 
$\hat x _{+} $. This is in marked contrast to schemes employing  feedback-assisted dynamics~\cite{Wilson2022,Diehl2022,Turkeshi2023PRL_AbsorbingState,Khemani2022}.

We first consider our scheme in the simple case $\hat H _{\mathrm{det}} =0$. Surprisingly, we find that using a single register mode $c$ (i.e.~setting $M=1$ in Eq.~\eqref{eq:qme.meas.ff.nmodes}) is sufficient to capture the desired entanglement dynamics. This can be seen by computing the entanglement of the total system state, with the bipartition between modes $a $ versus ($b$, $c$). As shown in Fig.~\ref{fig:mff_ent}, in the long-time limit $\gamma t \gg 1 $, the unconditioned state generated by the dissipative dynamics in Eq.~\eqref{eq:qme.meas.ff.nmodes} with $\eta=1$ and all the modes in initial vacuum states (solid curve) fully preserves the conditioned entanglement just from measuring $\hat x _{+} $ (dotted curve). 
We also provide analytic arguments in Ref.~\cite{SI} for why a single register mode is sufficient in this case, 
and show that the entanglement of formation 
$\mathcal{E} _{F} $ has similar dynamics to the log negativity.  
While we took the relative feedforward strength $\eta = 1$ here, using larger values always enhances our scheme's ability to capture entanglement, as it makes one less sensitive to quantum noise in the initial register state~\cite{SI}.  We will nonetheless focus throughout on modest $\eta$ values, as this is sufficient for good performance and much more compatible with experiment.

While Eq.~\eqref{eq:qme.meas.ff.nmodes}
generates entanglement that closely mimics the conditional dynamics of the original monitored system, it does not generate a pure state; in fact, the purity decreases with time
(see Ref.~\cite{SI} for an analogous effect in a qubit system).  Using the measurement-feedforward interpretation of Eq.~\eqref{eq:qme.meas.ff.nmodes}, this can be attributed to the
random nature of a given measurement trajectory.  
Fortunately, as shown in the dashed curves in Fig.~\ref{fig:mff_ent}, one can still recover almost pure-state entanglement from this highly mixed state via a single Gaussian, projection measurement on mode $c $, followed by local, conditional feedback operations on the target modes (see Ref.~\cite{SI}).

\begin{figure}[t]
    \centering
    \includegraphics[width=\columnwidth]{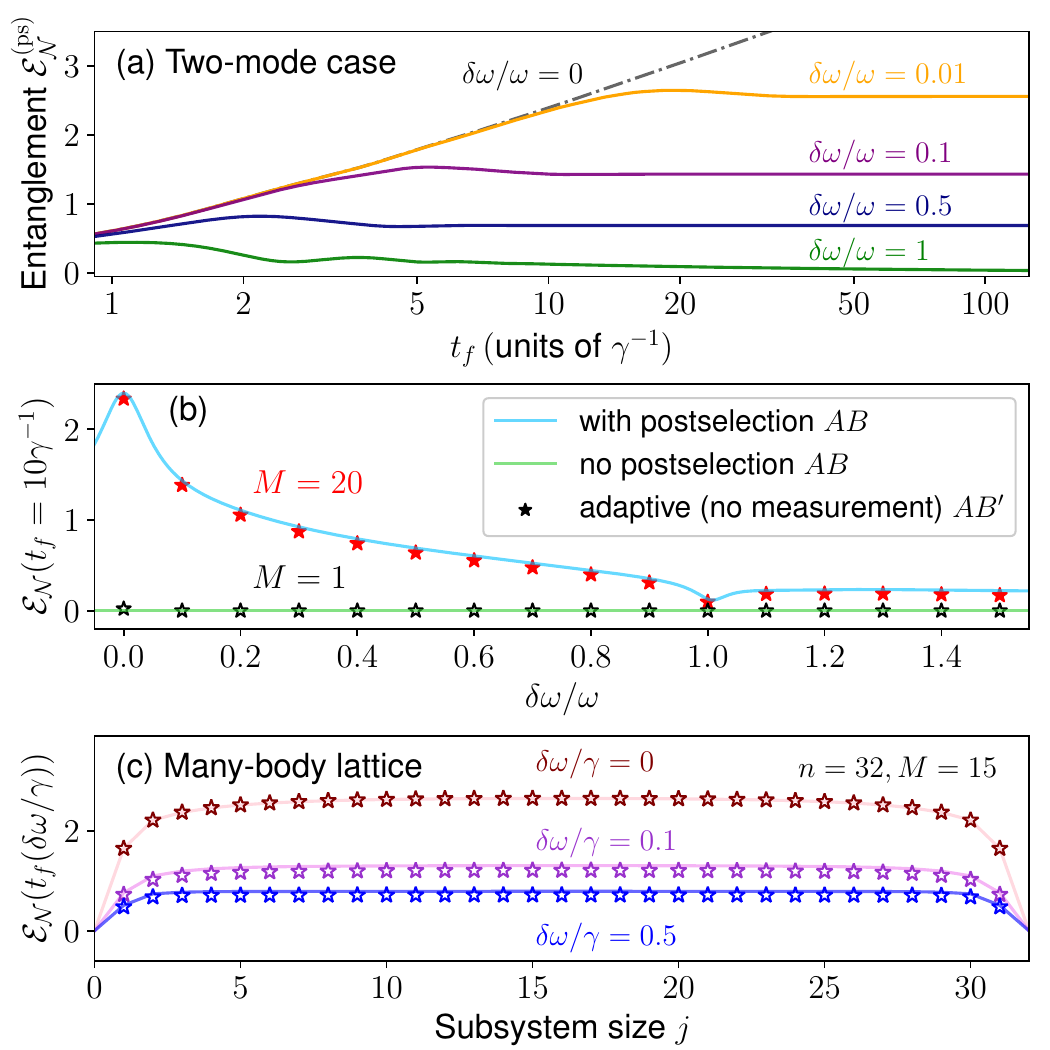}
    \caption{(a) Time-dependent $(a,b)$ entanglement in the postselected state generated from the combination of Hamiltonian dynamics as per 
    Eq.~\eqref{eq:h.qmfs.perturb}
    and continuous measurement of $\hat {x} _{+} $. (b) Light blue: measurement-induced entanglement at a fixed evolution time $t _{f}=10 \gamma^{-1}$, as a function of the Hamiltonian parameter $\delta \omega$.  Red asterisks:  same, but entanglement generated by our measurement-free adapative dynamics scheme with $ M =20 $ registers (see Eq.~\eqref{eq:qme.meas.ff.nmodes}), with respect to the bipartition between $ a $ and the rest of the system.  
    The adaptive scheme quantitatively captures the non-trivial parameter dependence of the conditional entanglement generated in the original measured setup. Parameters in (a) and (b): $\omega / \gamma = 1.2$, $\eta =5 $, $t _{f}=10 \gamma^{-1}$. (c) Solid lines: entanglement page curve (between the first $j$ sites and the rest of the system) generated by conditional dynamics under uniform onsite detunings ($\delta \omega \ne 0 $) and measurements on nearest neighbor bonds (with strength $\gamma$) on a $32$-site lattice; asterisks: entanglement created via the corresponding measurement-free adaptive protocol (see Eq.~\eqref{eq:qme.meas.ff.nmodes}). Parameters: 
    $M=15$, $\eta =5$. Total protocol time is chosen as $t_f =50\gamma ^{-1}$ for $\delta \omega = 0 $, or the time when entanglement stabilizes in the conditional dynamics or the measurement-free adaptive protocol, respectively  (see~\cite{SI} for details).  
    }
    \label{fig:qmfs.det}
\end{figure}

{\it Capturing non-trivial entanglement features-- }
We return to Eq.~\eqref{eq:qme.meas.ff.nmodes}, and now consider dynamics with both collective measurement and a competing local Hamiltonian.
Consider first features of the measured dynamics with postselection, which as mentioned can exhibit unbounded entanglement growth ($\delta \omega = 0$) or saturation ($\delta \omega \ne 0$) in the long-time limit, as shown in Fig.~\ref{fig:qmfs.det}(a). Surprisingly, adding a local Hamiltonian can enhance long-time entanglement growth (logarithmic growth that is twice as fast compared to Eq.~\eqref{eq:meas.cond.xp.ent})
when $\delta \omega = 0$ and $ \omega \ne 0$. As shown in Ref.~\cite{SI}, this can be understood via a special symmetry structure of $\hat H _{\mathrm{det}}$, often termed a  quantum-mechanics-free subsystem (QMFS)~\cite{Caves2012,Polzik2017}.  If the system is weakly perturbed by tuning $\delta \omega$ away from the QMFS parameter point 
the logarithmic negativity at a fixed evolution time exhibits a sharp peak at $\delta \omega = 0$ (see the light blue curve in Fig.~\ref{fig:qmfs.det}(b)). The entanglement of the conditional state also shows curious nonmonotonic-in-time features when the perturbation crosses the point $\delta \omega = \omega $ (c.f.~green curve in Fig.~\ref{fig:qmfs.det}(a)); at that specific parameter choice, the entanglement vanishes in the asymptotic long-time limit.

We now come to a central result of this work:  the nontrivial features in the postselected entanglement generation discussed above can be faithfully captured in the {\it unconditional} state produced by the dissipative dynamics of Eq.~\eqref{eq:qme.meas.ff.nmodes} (see Fig.~\ref{fig:schematic}(b)). We plot in Fig.~\ref{fig:qmfs.det}(b) the  entanglement generated between $a $ (subsystem $A$) and the expanded system of $ b $ and $M$ register modes (subsystem $B'$) as a function of $\delta \omega$ (obtained numerically from Eq.~\eqref{eq:qme.meas.ff.nmodes}).   
Using $M=20$ register modes (corresponding to a coarse-graining timescale $\Delta t \sim t_f/20 = 0.5 \gamma^{-1}$) allows us to closely reproduce the postselected entanglement dynamics, even though this retains a tiny fraction of the full measurement records~\footnote{Note that at the QMFS point, we only need $2$ auxiliary register modes to fully capture long-time conditional entanglement generation using a protocol that is similar to Eq.~\eqref{eq:qme.meas.ff.nmodes} but with different time-modulating functions $f_{j} (t; t_{f})$.} (see Ref.~\cite{SI} for more details).

{\it Generalizations-- }The adaptive-dynamics approach can be generalized to a variety of 
more complex systems.    
As a many-body extension of our bosonic example, we consider a $1$-dimensional bosonic lattice subject to continuously monitoring of linear bond variables. 
Such dynamics can create long range entanglement in the conditional state and host measurement-induced entanglement transitions~\cite{Buchhold2022,Ashida2024}. Further, the conditional entanglement transitions from log-law ($\delta \omega = 0$) to area-law ($\delta \omega \ne  0$) in the presence of an onsite Hamiltonian, $\hat H _{\mathrm{det,multi}} = 
\delta \omega \sum _{j=1} ^{n}  \hat a _{j}^{\dag}
\hat a _{j} $. As shown in Fig.~\ref{fig:qmfs.det}(c), a direct multimode generalization of the autonomous protocol in Eq.~\eqref{eq:qme.meas.ff.nmodes} lets us capture conditional entanglement generation in the two regimes without any measurement or postselection (see Ref.~\cite{SI} for simulation details). Importantly, the coarse graining procedure as per Eq.~\eqref{eq:qme.meas.ff.nmodes} quantitatively preserves spatial entanglement structure using values of the feedforward parameter ($\eta=5$) and number of registers per bond ($M=15$) that are comparable to the $2$-mode case. Further simulations with varying system sizes~\cite{SI} show that for our protocol to quantitatively reproduce the conditional entanglement generation, the required $M $ and $ \eta$ do not scale with $n$, indicating that our protocol generalizes to many-body systems with a hardware overhead scaling linearly in system size.

It is also worth noting that the bosonic systems discussed thus far can be directly mapped to genuinely many-body qubit systems, where each bosonic mode becomes an ensemble of $2S$ spins ($S \gg 1$) and the quadrature operator $ \hat x_{a}$ is replaced with the rescaled total angular momentum operator of $a$-th ensemble along $x$ axis, $ \hat x_{a} \to   \hat S_{a,x}/\sqrt{S} $, via the Holstein-Primakoff transformation~\cite{Holstein1940}.
The aforementioned bosonic $1$D system can thus describe an array of spin ensembles with nontrivial connectivity; such systems with programmable connectivity have been recently realized experimentally using atomic ensembles, see e.g.~\cite{SchleierSmith2023,SchleierSmith2024}.

\begin{figure}[t]
    \centering
    \includegraphics[width=\columnwidth]{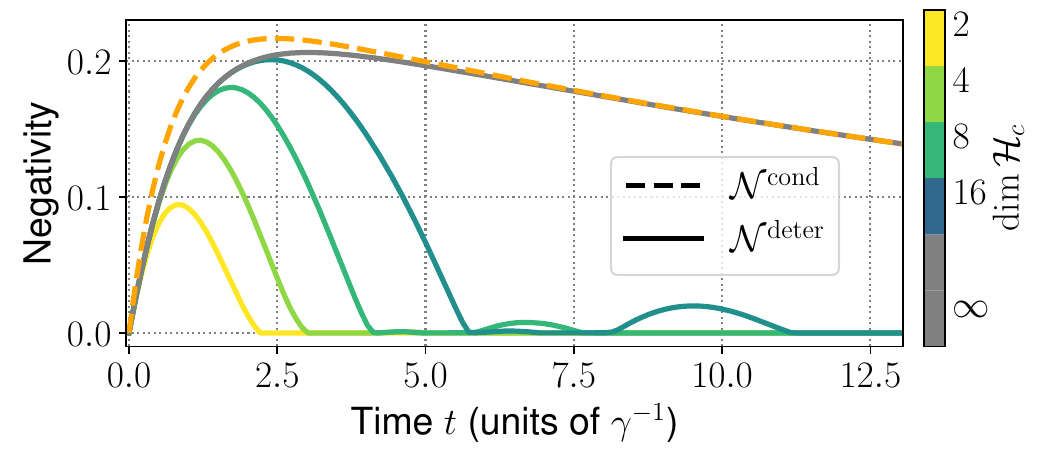}
    \caption{Deterministic entanglement generation between qubit $1$ and extended systems consisting of qubit $2$ and an auxiliary $d$-dimensional register $c$, via the adaptive dynamics in Eq.~\eqref{eq:qme.mff.2qb} that realize an autonomous measurement-and-feedforward process (solid curves). In the asymptotic long-time limit and for $c$ being a bosonic mode with a quadrature as the feedforward operator, the deterministic entanglement (gray solid curve) perfectly replicates the trajectory-averaged conditioned entanglement (black dashed curve) created by weak nonlocal measurement of $\hat \Sigma _{x} 
\equiv (\hat \sigma _{x, 1 }  
+ 0.7 \hat \sigma _{x, 2 }  )/ 2 $ at intermediate and large timescales. Parameter: $\eta =1$. 
    }
    \label{fig:qubit_mff_trunc_ho}
\end{figure}

Another generalization is to move beyond quadratic bosonic systems.  This immediately creates extra complexity, as now, measurement-induced entanglement in the postselected state can fluctuate from trajectory to trajectory~\footnote{This is similar to the situation of going from Clifford to non-Clifford circuits.}.  
Despite this, for several examples in this more general category, we find that our adaptive scheme still provides a means for replicating conditional measurement-induced entanglement.  
For concreteness, consider a target system of two qubits undergoing a continuous measurement of a nonlocal operator $\hat \Sigma_x$.  For our autonomous scheme, we will use a single $d$-dimensional auxiliary register system, yielding a dynamics    
\begin{align}
\label{eq:qme.mff.2qb}
&  (d \hat \rho  / \gamma dt) =
- i 
[\eta \hat \Sigma _{x}  
\hat F , 
\hat{ \rho} ] 
+  
\mathcal{D} [ 
\hat \Sigma _{x}  
- i \eta \hat F   ] 
\hat{ \rho}  
, 
\end{align}
where 
$\eta$, $\hat F$ denote the feedforward strength and feedforward operator acting on the  register, respectively. We take $\hat \Sigma _{x} 
\equiv (\hat \sigma _{x, 1 }  
+ 0.7 \hat \sigma _{x, 2 }  )/ 2 $ 
and $\hat F $ a truncated bosonic quadrature operator acting on the auxiliary qudit (note the applicability of our protocol does not rely on specific form of these operators, see Ref.~\cite{SI}). As shown in Fig.~\ref{fig:qubit_mff_trunc_ho}, the measurement-free dynamics in Eq.~\eqref{eq:qme.mff.2qb} accurately captures features of the conditioned entanglement  at short to moderate times.  At longer times there is no longer a correspondence, something that is simply understood as being the result of the finite dimensionality of the register (e.g.~there is no longer sufficient dynamic range to store measurement outcomes).  Increasing dimension of the register $d \equiv \dim \mathcal{H} _{c}$ (see Fig.~\ref{fig:qubit_mff_trunc_ho}) increases the effective dynamic range of the register, thus enhancing its ability to store information.

{\it Discussions-- } We have presented a general measurement-free method for replicating conditional entanglement generated in a continuously monitored system. 
It involves engineering directional interactions with an auxiliary register system which effectively retains a small fraction of the information content in a given measurement record.  
The examples we consider suggest 
that despite the seemingly large loss of information contained in the specific measurement outcomes, this approach can quantitatively capture entanglement features arising from the competition between measurements and unitary dynamics, as measured by negativity-based quantities. 
In future work, it would be interesting to understand more rigorously the question of how much of the measurement record in a generic setup can be discarded before our scheme breaks down. 
While our work focused on logarithmic negativity, another question worth further exploration is whether one can recover more general entanglement measures, such as higher R\'{e}nyi entanglement entropies, via our measurement-free protocol.

We note that the key ingredients of our scheme can be realized either directly in physical platforms that inherently allow nonreciprocal QND interactions, e.g.,~in levitated nanoparticles~\cite{Rieser2022}, or synthetically using standard reservoir engineering techniques~\cite{Zoller1996,Murch2022}; the latter has been demonstrated in a variety of experimental platforms~\cite{Polzik2011,Home2015,Home2019,Schwab2015,Sillanpaa2015,Schwab2016} (see Ref.~\cite{SI} for detail).

Our work ultimately employs feedforward dynamics, and hence connects to other general topics in quantum dynamics.  This includes ideas for how feedback and feedforword can enhance the power of shallow quantum circuits~\cite{Ignacio2021,Verresen2023,Minev2023}.  It also connects to the question of if and when dissipative Markovian dephasing dynamics can create entanglement, an issue that can also be mapped to effective feedforward processes~\cite{Seif2022}.
Our results here provide an intuitive understanding of how entanglement can be generated by such correlated dephasing processes (see Ref.~\cite{SI}).

{\it Acknowledgments-- }  We thank Marko Cetina, Liang Jiang, and Michael Gullans for useful discussions.
This work was supported by the Air Force Office of Scientific Research under Grant No. FA9550-19-1-0362.  
A. C. also acknowledges support from the Simons Foundation through a Simons Investigator Award (Grant No. 669487, A. C.).

%


\clearpage

\onecolumngrid
\begin{center}
\textbf{\large Supplementary Material: Uncovering measurement-induced entanglement via directional adaptive dynamics and incomplete information}
\end{center}

\setcounter{equation}{0}
\setcounter{figure}{0}
\setcounter{page}{1}
\renewcommand{\theequation}{S\arabic{equation}}
\renewcommand{\thefigure}{S\arabic{figure}}

\vspace{.5cm}
\twocolumngrid

\tableofcontents

\subsection{Overview of content}

Here we provide derivation details related to various examples discussed in the main text, intuition for increased entanglement generation accompanying decreasing purity in the adaptive protocol, as well as more technical details for experimental implementations.

\subsection{Postselected trajectory dynamics due to measurement of collective quadrature $\hat x _{+} $}

\label{suppsec:cond.dyn.gaus}

\subsubsection{General trajectory dynamics}

In the main text, we discuss conditional entanglement generation due to the weak continuous measurement of a nonlocal quadrature 
$\hat {x} _{+} \equiv 
\left (\hat {x} _{a} + 
\hat {x} _{b} \right) / \sqrt{2} $. Here, we provide a detailed analysis on the conditional dynamics generated by this process (in the absence of any additional Hamiltonian dynamics). 

Given the measurement current $I(t) $ associated with a specific run of the experiment, we can explicitly derive the corresponding evolution of the conditional state of the systems density matrix.  This is written in terms of the measurement rate 
$\gamma$ and a stochastic Wiener increment 
$dW (t)$, as (see e.g.~\cite{Steck2006} for a pedagogical introduction): 
\begin{align}
\label{eq:sqme.meas.xp}
& d \hat \rho = 
- \frac{\gamma dt }{2} 
\left[ 
\hat x _{+} , \left[ 
\hat x _{+} , \hat{ \rho} 
	\right ]
	\right ]  
	+ \sqrt {\gamma} d W 
\left\{ \hat x _{+} - 
\left \langle \hat x _{+}
\right \rangle , \hat{ \rho}  \right \} 
. 
\end{align}
The measurement record $I(t) $ is in turn updated as:
\begin{align}
\label{seq:sto.record.xp}
dI(t) = 2 \sqrt{\gamma} 
\langle \hat x _{+} (t) \rangle 
dt + dW (t). 
\end{align}
A qubit version of this dynamics was discussed in Ref.~\cite{Korotkov2003}.

Throughout this work, we consider initial states that are Gaussian.  The above dynamics then ensures that the conditional state remains Gaussian at all times, and can thus be fully characterized by its first $2$ moments.  We will further assume a state where there are no correlations between the $+$ and $-$ sets of quadratures.  In this case, the relevant dynamical quantities are the means 
$\langle \hat x _{+} \rangle$, 
$\langle \hat p _{+} \rangle$, and variances 
$ V   _{\sigma_1\sigma_2}
\equiv \langle \{ 
\delta (\hat \sigma_1 )  _{+} , 
\delta ( \hat \sigma_2 ) _{+} 
\} \rangle
/2$. The conditional dynamics can thus be described by the equations of motion as follows 
\begin{subequations}
\label{eq:meas.cond.eom}
\begin{align}
\label{eq:meas.cond.eom.xp}
& d \langle \hat x _{+} \rangle
= 2 \sqrt{\gamma} V_{xx} dW
, \,
d \langle \hat p _{+} \rangle
= 2 \sqrt{\gamma} V_{xp} dW
, \\ 
\label{eq:meas.cond.eom.vxx}
& d V_{xx} 
= -4 \gamma 
V_{xx} ^2 
dt
, \, 
d V_{pp} 
= \gamma dt
-4 \gamma 
V_{xp} ^2 
dt 
, \\
\label{eq:meas.cond.eom.vxp}
& d V_{xp} 
= -4 \gamma 
V_{xx} V_{xp} 
dt
.   
\end{align}
\end{subequations}
Note that the dynamics of the covariance matrix, which fully determines entanglement properties of the state, is deterministic because the quadrature measurement is Gaussian. This makes the postselected system dynamics particularly simple to characterize.

Physically, Eqs.~\eqref{eq:meas.cond.eom.vxx} and~\eqref{eq:meas.cond.eom.vxp} describe measurement-induced squeezing of the collective quadrature 
$\hat x _{+} $. As a result, the conditional state is entangled between modes 
$a $ and $ b $. If the system starts in the vacuum as its initial state, we can compute entanglement entropy of the conditional state $ \mathcal{S}  
^{\mathrm{(ps)}} $, as 
\begin{align}
& \mathcal{S}  
^{\mathrm{(ps)}}
=  \frac{\nu _{t} +1 }{2}
\log  \frac{\nu _{t} +1 }{2}
- \frac{\nu _{t} - 1 }{2}
\log 
    \frac{\nu _{t} - 1 }{2}
    , 
\end{align}
where $\nu _{t}
\equiv (1+  \gamma t) / 
{\sqrt {1+ 2
\gamma t  } }$ denotes the symplectic eigenvalue of the mode $a$ covariance matrix. It is interesting to note that in the long-time limit, the conditional entanglement grows logarithmically: 
$\mathcal{S}  
^{\mathrm{(ps)}} 
\sim (1/2)\log ( \gamma t) $. As the conditional state is pure, one can show that this logarithmic growth is also seen if one uses other entropic entanglement measures, or the logarithmic negativity.  This slow but unbounded entanglement growth is reminiscent of entanglement growth in a critical many-body system.  Here, it can be understood directly as resulting from the squeezing of the EPR variable $\hat{x}_+$ that results from the continuous measurement.

We stress that the entanglement discussed above only exists in the conditional state within the stochastic evolution. Conversely, if we average over all measurement records, the unconditional evolution cannot generate any entanglement. This agrees with the picture provided by the stochastic equation of motion Eq.~\eqref{eq:sqme.meas.xp}: averaging over the random variable $d W $, we obtain a Lindblad quantum master equation  
\begin{align}
\label{eq:qme.xp}
& \frac{d \hat \rho }{dt} = 
- \frac{\gamma}{2}  
\left[ 
\hat x _{+} , \left[ 
\hat x _{+} , \hat{ \rho}  
	\right ]
	\right ] 
=\gamma  
\mathcal{D}\left[ 
\hat x _{+}  \right ] 
\hat{ \rho}  
, 
\end{align}
where the standard Lindblad dissipator is defined as $  \mathcal{D} [\hat{O}  ]  
\hat{\rho} \equiv  
 (\hat{O}  \hat{\rho} \hat{O} ^\dagger 
- \{\hat{O} ^\dagger \hat{O}
, \hat{\rho}\} /2  ) $. As Eq.~\eqref{eq:qme.xp} also describes system dynamics that would arise from \textit{locally} driving the two modes with the same noisy classical force, it cannot generate entanglement between $ a $ and $ b $. 

\subsubsection{Proof of exact equivalence between coarse grained dynamics and conditional trajectories under a single commuting measurement}

Making use of the equations of motion of the conditional system dynamics Eq.~\eqref{eq:meas.cond.eom}, we can also formally solve the dynamics of the stochastic conditional displacement $ \langle \hat x _{+} (t) \rangle $, as well as the measurement current $ I(t) $. Integrating Eqs.~\eqref{seq:sto.record.xp} and~\eqref{eq:meas.cond.eom.xp} explicitly, we have 
\begin{align}
& \langle \hat x _{+} (t) \rangle
= 2 \sqrt{\gamma} \int ^{t} _{0}
V_{xx} (t') dW (t')
, \\ 
&  I(t) = 2 \sqrt{\gamma} 
\int ^{t} _{0} 
\langle \hat x _{+} (t') \rangle 
dt' + \int ^{t} _{0} dW (t'). 
\end{align}
If the initial state is vacuum, the covariance matrix dynamics is exactly solvable, so that we obtain
\begin{align}
& d \langle \hat x_+ (t) \rangle
= \sqrt{\gamma} 
\frac{ dW (t)}{ 1 + 2 \gamma t}
, \\ 
& dI(t)  = 2 \sqrt{\gamma} 
\langle \hat x_+   (t) \rangle 
dt + dW (t). 
\end{align}
We can further rewrite those equations as integrals, i.e.
\begin{align}
&\langle \hat x_+ (t  ) \rangle 
=   \sqrt{\gamma} \int ^{t  } _{0}
\frac{ dW (t _{1})}{ 1 + 2 \gamma t _{1} }
, 
\\
\label{seq:meas.record.int.dW}
& I(t  ) = 2 \gamma \int ^{t _{f} } _{0} 
d t _{2} 
\int ^{t _{2}  } _{0}
\frac{ dW (t _{1})}{ 1 + 2 \gamma t _{1} }
+  \int ^{t  } _{0}
dW (t'). 
\end{align}
We can now rewrite the measurement record $I(t  ) $ Eq.~\eqref{seq:meas.record.int.dW} in terms of the expectation value of the quadrature operator $\langle \hat x _{+} (t ) \rangle $, as 
\begin{align}
I(t ) = & 2 \gamma \int ^{t } _{0} 
\frac{ dW (t _{1})}{ 1 + 2 \gamma t _{1} }
\int ^{t } _{t _{1} } d t _{2} 
+  \int ^{t } _{0}
dW (t')
\nonumber \\
=& \int ^{t  } _{0} 
\left ( \frac{2 \gamma  t -2 \gamma t _{1}}{ 1 + 2 \gamma t _{1} }
+ 1 \right )
dW (t _{1})
\nonumber \\
=& \int ^{t  } _{0} 
 \frac{ 1 + 2 \gamma  t  }{ 1 + 2 \gamma t _{1} }
dW (t _{1})
\nonumber \\
= & \frac{ 1 + 2 \gamma  t  }{ \sqrt{\gamma}  }
\langle \hat x_+ (t   ) \rangle 
. 
\end{align}

This results tells us that knowledge of the integrated measurement record $I(t)$ is all that is needed to reconstruct the stochastic displacement $\langle \hat{x}_+(t) \rangle$, the only stochastic parameter in the full description of the conditional state.  This in turn tells us why the autonomous protocol using $M=1$ modes in the main text becomes perfect for large feedforward strength:  the only information needed to capture the stochastic part of the conditional state is the integrated measurement record.

\begin{figure}[t]
    \centering
    \includegraphics[width=\columnwidth]{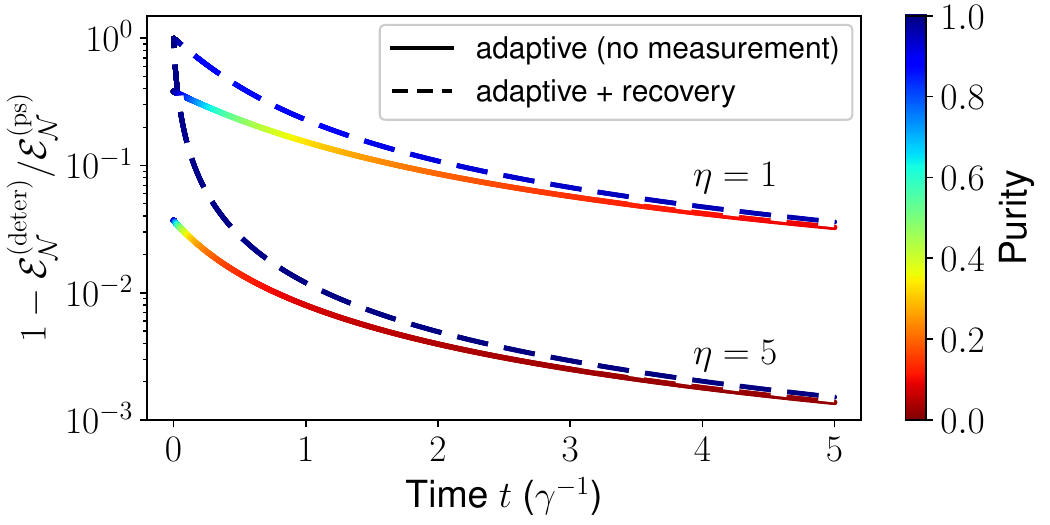}
    \caption{Efficiency of preserving postselected entanglement generation via adaptive dynamics. Plotted is entanglement inefficiency, defined as the relative difference between the deterministic state entanglement 
    $\mathcal{E} _{\mathcal{N}} ^{\mathrm{(deter)}}$ (between modes 
    $a $ and $(b, c) $), and the entanglement of the postselected states $\mathcal{E} _{\mathcal{N}} ^{\mathrm{(ps)}}$ (between modes $a$ and $b$) as generated by continuous measurement of the quadrature $\hat {x} _{+} \equiv 
\left (\hat {x} _{a} + 
\hat {x} _{b} \right) / \sqrt{2} $ (see Eq.~\eqref{eq:meas.cond.xp.ent}). 
    Colors of the curves denote purity of the state. Solid curves correspond to states generated by measurement-free dissipative (adaptive) dynamics in Eq.~\eqref{eq:qme.meas.ff}, whereas dashed curves represent the states after adaptive dynamics and a recovery via projection measurement of mode $c$ quadrature 
$\hat \pi $ and conditional feedback. 
    In both cases, the entanglement monotonically increases for stronger feedforward strength $\eta$.   
    The state purity before recovery degrades for increasing $\eta$; in contrast, the purity of post-recovery state increases for greater $\eta$.  }
    \label{suppfig:mff_ent}
\end{figure}

\subsection{Parameter dependence and entanglement structure of states generated by the adaptive process Eq.~\eqref{eq:qme.meas.ff}}

\label{suppsec:eof}

In the main text, we show that the measurement-free dynamics in Eq.~\eqref{eq:qme.meas.ff} can be used to deterministically create entanglement that quantitatively preserves the entanglement generation in Eq.~\eqref{eq:meas.cond.xp.ent} due to measurement-induced postselected dynamics. To quantify the efficiency of this procedure, we compute the entanglement inefficiency, i.e., the fraction of conditional entanglement that is missed when applying the dissipative dynamics in Eq.~\eqref{eq:qme.meas.ff}. As depicted by the solid curves in Fig.~\ref{suppfig:mff_ent}, this inefficiency monotonically decreases as we increase the feedforward parameter $\eta$ (see Eq.~\eqref{eq:qme.meas.ff}). Furthermore, the unconditioned state fully preserves the measurement-induced entanglement growth with postselection in the long-time limit $\gamma t \gg 1 $, regardless of the specific value of $\eta$.

\begin{figure}[t]
    \centering
    \includegraphics[width=\columnwidth]{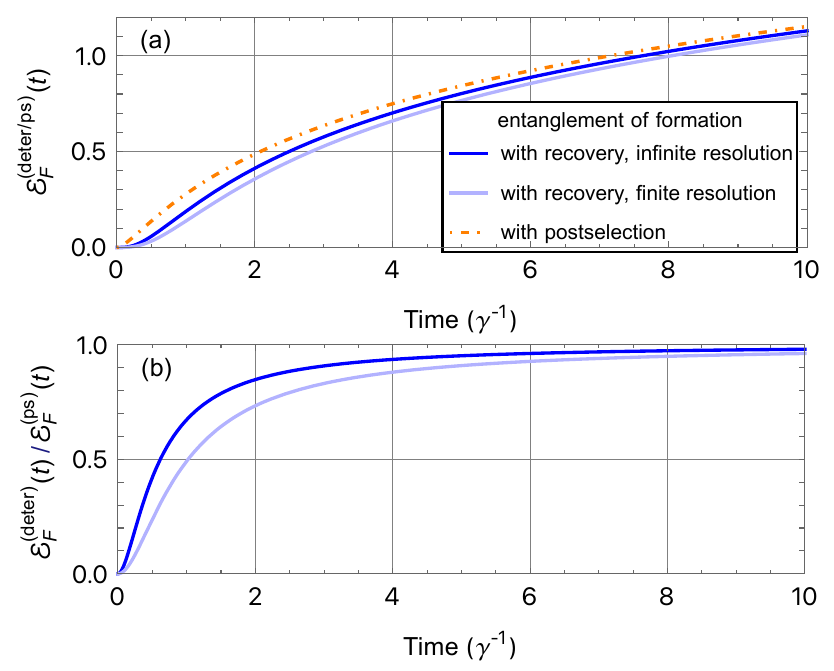}
    \caption{(a) Deterministic entanglement of formation $\mathcal{E} _{F} ^{\mathrm{(deter)}} $ for $2$-mode states obtained from recovery via projective measurement of $\hat \pi$ and local feedback conditioned on measurement result. The case of entanglement generation from measuring $\hat x _{+} $ with postselection $\mathcal{E} _{F} ^{\mathrm{(ps)}} $ is also plotted as a reference (orange dot-dashed curve, computed from Eq.~\eqref{eq:meas.cond.eom}).  
    (b) The ratio between entanglement of formation $\mathcal{E} _{F} $ of deterministic and postselected states. In the long-time limit, the protocol with recovery can asymptotically fully restore measurement-induced conditional entanglement; as discussed in the main text, this state is also very close to a pure state (see also Fig.~\ref{suppfig:mff_ent}).}
    \label{suppfig:sm_eof_ent}
\end{figure}

Noting that the unconditional state evolving under Eq.~\eqref{eq:qme.meas.ff} is generally mixed, its entanglement can be complicated to characterize. While logarithmic negativity $\mathcal{E} _{\mathcal{N}}$ provides an easy-to-compute method to verify that the state is entangled, its absolute value only offers an upper bound on the distillable entanglement. From the perspective of quantum information processing, it would be valuable to be able to characterize genuine entanglement measures (e.g.,~entanglement of formation $\mathcal{E} _{ F }$) of our final state.
However, it is very challenging to exactly compute such entanglement measures for a generic $3$-mode mixed Gaussian state. Fortunately, making use of the basic property that entanglement measures cannot increase under any local operation and classical communication (LOCC) processes, we can compute rigorous lower bounds for a generic entanglement measure that we want to characterize. More specifically, the deterministic state after recovery operation discussed in the main text and App.~\ref{suppsec:rec} offers an example of such ``diagnosing" states (c.f.~Eq.~\eqref{seq:gaus.meas.fb}). 

In Fig.~\ref{suppfig:sm_eof_ent}, we plot the numerically computed entanglement of formation for the unconditional $2$-mode state after recovery, which provides a lower bound for the entanglement of formation of the full $3$-mode state generated from the adaptive protocol Eq.~\eqref{eq:qme.meas.ff}, along with the conditional entanglement generated from continuously measuring $\hat x _{+} $. It is clear that with sufficiently long evolution time, our protocol can fully recover the measurement-induced conditional entanglement. Combined with the purity of the recovered states shown in Figs.~\ref{fig:mff_ent} and~\ref{fig:pur_ent}, we see that in the asymptotic long-time limit, the recovery operation can convert the mixed-state entanglement from the measurement-and-feedforward protocol into almost pure entangled states with perfect efficiency.

\subsection{Intuition for higher entanglement generation accompanying decreasing purity in time under the adaptive protocol}

\label{suppsec:qb.mixed.ent}

In the main text, we show that the state generated by the adaptive dynamics in Eq.~\eqref{eq:qme.meas.ff.nmodes} sees its purity decreases as time grows and entanglement generation increases.  
To better understand the (mixed state) entanglement generation mechanism in this case, it is useful to consider a discrete version of our measurement-and-feedforward protocol. For simplicity we focus on a $3$-qubit system, which starts in a product initial state 
$\left | \downarrow\downarrow\downarrow \right\rangle$, where 
$\left | \downarrow \right\rangle$ is eigenstate of the Pauli operator with 
$\hat \sigma _{z } = 
\left |\uparrow \rangle 
\langle \uparrow \right |
- \left |\downarrow \rangle 
\langle \downarrow \right | $. As we discuss, this setup can be viewed as the qubit-analog of the bosonic system in Eq.~\eqref{eq:qme.meas.ff}.

We now consider a process that can be viewed as the discretized version of the measurement-and-feedforward dynamics. For simplification, we take the nonlocal measurement to be a strong projective measurement of the collective spin operator 
$ \hat \sigma _{x,1} 
+ \hat \sigma _{x,2} $ between the first two qubits, but we stress that the resulting mixed-state entanglement structure applies to weak and strong measurements alike. Contingent on the measurement result, the qubit system is projected into one of the $2$ Bell states $| \Phi _{\pm} \rangle$ with equal probabilities, where we have 
($ \hat \sigma _{x} =
\left | + \rangle 
\langle + \right |
- \left |- \rangle 
\langle - \right | $)
\begin{align}
& | \Phi _{+} \rangle
= (\left | ++ \right\rangle
+ \left | -- \right\rangle
)/\sqrt{2}
, \\
& | \Phi _{-} \rangle
= (\left | +- \right\rangle
+ \left | -+ \right\rangle
)/\sqrt{2}
. 
\end{align}
For convenience, we also define the corresponding density matrices as 
\begin{align}
\hat \rho _{\pm} 
= | \Phi _{\pm} \rangle
\langle \Phi _{\pm} | 
. 
\end{align}
At this stage, the averaged state 
$(\hat \rho _{+} + \hat \rho _{-} )/2$ has no entanglement between the two qubits. To preserve entanglement in the unconditional state, we construct a (classically) correlated state between the first $2$ qubits and the third qubit, in such a way that the classical information about the two Bell pairs is stored into the quantum state of the third qubit. This is again the discrete version of the feedforward operation. We thus obtain the following final state
\begin{align}
\label{seq:rhof.qb}
\hat \rho _{ \mathrm{qb} }
= \frac{1}{2}
( \hat  \rho _{+}
\otimes 
\left | \uparrow \rangle
\langle \uparrow \right |
+ 
\hat  \rho _{-}
\otimes 
\left| \downarrow \rangle
\langle \downarrow \right | ) 
. 
\end{align}
Similar to the bosonic case, there is only classical correlation between the first two and the third qubit. However, the feedforward process now ``protects" entanglement in the conditional $2$-qubit state against averaging over measurement results, so that we obtain unconditional entanglement between the first and the last two qubits in 
$\hat \rho _{ \mathrm{qb} }$.

The entanglement structure of the qubit state Eq.~\eqref{seq:rhof.qb} is straightforward to analyze. In fact, one can show that the qubit state 
$\hat \rho _{ \mathrm{qb} }$ can be converted to a standard $2$-qubit Bell pair (tensor product some third qubit state) solely using LOCC operations. 
As a result, the entanglement of formation $\mathcal{E} _{ F }$ and distillable entanglement $\mathcal{E} _{D}$ of $\hat \rho _{ \mathrm{qb} }$ are both equal to the entanglement entropy of a single Bell pair, i.e. 
\begin{align}
\label{seq:ent.qb}
\mathcal{E} _{F}
( \hat \rho _{ \mathrm{qb} } ) 
= \mathcal{E} _{D}
( \hat \rho _{ \mathrm{qb} } )
= \mathcal{S}
[ \text{Tr} _{1} (\hat \rho _{+/-} ) ] = 1 
. 
\end{align}
Here, we define $\mathcal{S} ( \hat \rho )$ as the von Neumann entropy of the (reduced) density matrix $\hat \rho$.

\subsection{Details of the recovery protocol}

\label{suppsec:rec}

\subsubsection{Summary}

In the main text, we state that one can use a local projection measurement on the auxiliary mode $ c $ and subsequent local unitary operations to unconditionally recover almost-pure entangled state between modes $a$ and $b$. Here, we provide a detailed discussion on this recovery procedure, as well as an explicit derivation of the state evolution under the recovery operation.  

While the measurement-and-feedforward process in Eq.~\eqref{eq:qme.meas.ff} (i.e.,~Eq.~\eqref{seq:qme.mff.gen}) can convert measurement-induced conditional entanglement into unconditional entanglement with almost perfect efficiency, the resulting state is highly mixed. As we show, a straightforward procedure involving a single Gaussian, projection measurement on mode $c $, followed by conditional feedback in the form of local displacements on the $a,b$ modes, could convert the mixed-state entanglement into near pure-state entanglement. Thus, we refer to it as the {\emph{recovery}} step hereafter.

\begin{figure}[t]
    \centering
    \includegraphics[width=\columnwidth]{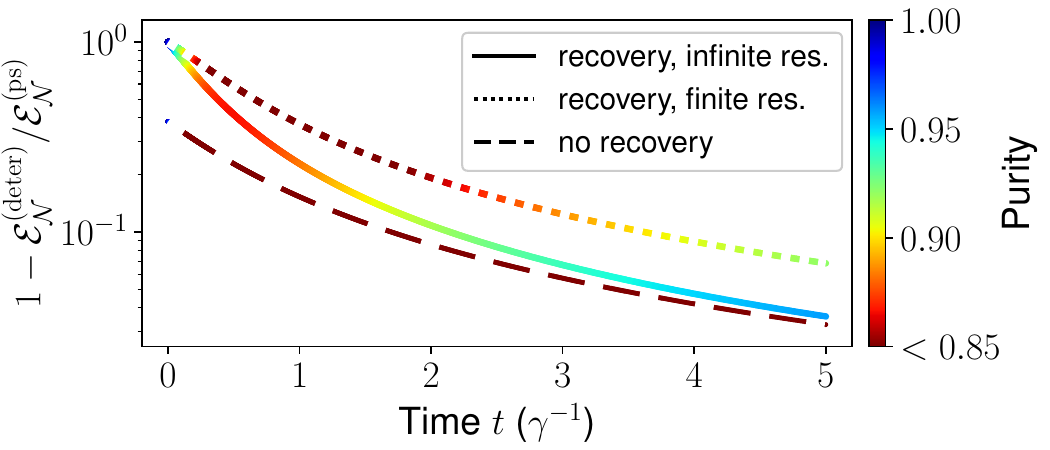}
    \caption{Entanglement inefficiency (defined in the same way as Fig.~\ref{fig:mff_ent}, i.e.~the ratio between the difference of the postselected and deterministic states entanglement 
    $ \mathcal{E} _{\mathcal{N}} ^{\mathrm{(ps)}} - \mathcal{E} _{\mathcal{N}} ^{\mathrm{(deter)}}$ and the postselected state value) and purity of states after recovery (see Eqs.~\eqref{seq:gaus.meas.povm} and~\eqref{seq:gaus.meas.fb}) for projection measurement of $\hat p _{3}$ with perfect resolution ($\mu \to \infty $, solid curve) versus finite resolution ($\mu =1 $, dotted curve). Other parameter: $\eta =1 $. 
    }
    \label{fig:pur_ent}
\end{figure}

We consider Gaussian quantum-non-demolition (QND) measurements of mode $c $ quadrature $\hat \pi $. The measurement resolution is quantified by $\mu$, which describes the variance of target quadrature operator in the measurement basis states (see the next subsection for an explicit definition). 
The performance of the recovery step then crucially depends on the feedforward strength $\eta $ and the measurement resolution $\mu $, and we generally expect the recovered entanglement to be upper bounded by the unconditional case generated by Eq.~\eqref{eq:qme.meas.ff}. We first consider the ideal case with ideal measurement resolution 
($\mu\to 0$), which corresponds to projection into
$ \hat \pi $ eigenstates, and numerically compute the entanglement between modes $a$ and $ b $, as well as purity of the recovered state. The results are shown in dashed curves in Fig.~\ref{suppfig:mff_ent}, in terms of the entanglement inefficiency, for different feedforward strengths. Interestingly, the recovery step can fully recover the original conditional entanglement in the long-time limit, and at the same time greatly improve purity of the state (see Fig.~\ref{suppfig:mff_ent}). The purity of the recovered state also increases for growing $\eta$ when fixing other parameters, in contrast to the case without recovery (see dashed curves in Fig.~\ref{suppfig:mff_ent}).

It is natural to ask if such high entanglement recovery efficiency survives under finite measurement resolution. As shown in Fig.~\ref{fig:pur_ent}, the near unity entanglement efficiency can be achieved in the asymptotic long-time limit, regardless of the intrinsic measurement uncertainty ($\mu$). In fact, one can rigorously show that the entanglement negativity 
$\mathcal{E} _{\mathcal{N}}  $ of the recovered state $\hat \rho 
_{\mathrm{rec} } 
\! \left ( t; \mu \right ) $ (see Eq.~\eqref{seq:gaus.meas.fb} for a definition) grows logarithmically in the asymptotic long-time 
$t\to \infty$ limit, i.e., 
\begin{align}
\label{eq:ent.pur}
t \to \infty: 
\, 
\mathcal{E} _{\mathcal{N}} 
( \hat \rho 
_{\mathrm{rec} } 
\! \left ( t; \mu \right ) 
 ) 
\sim \frac{1 }{ 2 } 
\log \left ( 
\gamma t \right)
+ o \left ( 
1 \right)
. 
\end{align}
Note that the leading-order contribution in Eq.~\eqref{eq:ent.pur} is independent of both $\eta$ and $\mu $. In this long-time limit, the unconditional state also approaches a perfectly pure state, see the dotted curve in Fig.~\ref{fig:pur_ent}.

\subsubsection{General setup of the recovery protocol}

We start by considering the Gaussian measurement on mode $c  $, which can be generally described by a POVM (positive operator-valued measure). The conditional states 
$\hat \rho 
_{\text{m} }
\! \left ( \zeta ; \mu \right )  $ corresponding to measurement outcomes $\zeta$ can be expressed using the POVM operators  $\hat{M}  _{\zeta, \mu} $ acting on $c$ as: 
\begin{align}
\label{seq:gaus.meas.povm}
&\hat \rho 
_{\text{m}}
\! \left ( \zeta ; \mu \right ) 
= \frac{( \hat{\mathbb{I} } _{ ab }
\! \otimes
\hat{M}  _{\zeta, \mu} )
\hat \rho 
(\hat{\mathbb{I} } _{ ab }
\! \otimes\hat{M}  _{\zeta; \mu} ^{\dag} ) 
}
{p  _{\text{m}}
\! \left ( \zeta ; \mu \right )  }
,  
\end{align}
where the parameter $\mu$ captures intrinsic resolution of the measurement. The measurement probability can be computed using the reduced density matrix 
$\hat \rho _{ c } \equiv 
\text{Tr} _{ ab } 
\hat \rho $, as 
$ p  _{\text{m}}
 ( \zeta ; \mu   ) = 
\text{Tr} ( \hat{M}  _{\zeta, \mu} ^{\dag}  
\hat{M}  _{\zeta, \mu} 
\hat \rho _{ c } ) $. In what follows, we focus on the case of projective Gaussian measurements on mode $c $ quadrature (and $\mu $ is directly set by variance of the measured quadrature in the states corresponding to projectors $\hat{M}  _{\zeta, \mu} $), but the formalism is straightforwardly applicable to other Gaussian measurements as well. The POVM operators can be explicitly written as mode $c $ projectors (see e.g.,~\cite{Plenio2002,Mista2007})
\begin{align}
\label{eq:gaus.meas.op}
& \hat{M}  _{\zeta, \mu} 
= \frac{ \hat{D}  _{c} 
\! \left ( \zeta  \right )
\hat{S}  _{c} 
\! \left (  \ln \mu  \right )
\left | 0 \rangle _{c} 
\langle 0 \right | 
\hat{S}  _{c} ^{\dag} 
\! \left (  \ln \mu  \right )
\hat{D} _{c} ^{\dag}
\! \left ( \zeta \right )  
}
{ \sqrt{2\pi} } 
, 
\end{align}
where $\left | 0 \right \rangle_{c} $ denotes mode 
$ c $ vacuum state, and we define the displacement and squeezing operators on mode $c $ as $\hat{D}  _{c} 
\! \left ( \zeta \right )
\equiv 
e ^{\zeta  
\hat c ^{\dag}  
- \zeta ^{*} 
\hat c } $ and $\hat{S}  _{c} 
\! \left ( r \right ) \equiv 
e ^{\frac{r}{2} ( \hat c ^{\dag} ) ^{2}
- \frac{r}{2} \hat c ^{2} } $, respectively. The POVM operators satisfy normalization condition 
$\int 
\hat{M} _{\zeta, \mu} ^{\dag} 
\hat{M}  _{\zeta, \mu}
d^{2} \zeta  
= \hat {\mathbb{I} } _{ c } $. 
Physically, Eq.~\eqref{eq:gaus.meas.op} can describe projection onto eigenstates of quadrature 
$\hat \pi $  or $\hat y $ ($\mu \to 0$ or $\mu \to +\infty $), or measurement of $\hat \pi $ with an intrinsic uncertainty set by $\mu$. Such measurements can be readily implemented in state-of-the-art optical or mechanical quantum systems.

Gaussian nature of the strong measurement in Eq.~\eqref{seq:gaus.meas.povm} ensures that the conditional states all have the same covariance matrix and only differ in terms of quadrature averages 
$ \langle \hat x _{\ell} \rangle$, 
$ \langle \hat p _{\ell} \rangle$, $ \langle \hat y \rangle$ and $ \langle \hat \pi \rangle$. We can thus apply conditional displacements 
$\hat{D} _{\mathrm{tot} }  
\!  \left ( \boldsymbol{\zeta} \right )$ after measuring $c $ to recover entanglement in the deterministic state. One can show that the optimal conditional feedback reproduces the same entanglement structure as the conditional state in Eq.~\eqref{seq:gaus.meas.povm}. The corresponding recovered state can be formally described as
\begin{align}
\label{seq:gaus.meas.fb}
\hat \rho 
_{\mathrm{rec} } 
\! \left (  \mu \right ) 
\equiv & 
\sum _{\zeta} 
p  _{\text{m}}
\! \left ( \zeta ; \mu \right ) 
\hat{D} _{\mathrm{tot} }  ^{\dag} 
\! \left ( \boldsymbol{\zeta} \right )
\hat \rho 
_{\text{m}}
\! \left ( \zeta ; \mu \right ) 
\hat{D} _{\mathrm{tot} }  
\! \left ( \boldsymbol{\zeta} \right ) 
\nonumber \\
= & \hat \rho 
_{\text{m}}
\! \left ( \zeta =0 ; \mu \right ) 
.   
\end{align}
See App.~\ref{suppsec:rec.dyn} for details and an explicit expression for 
$ \hat{D} _{\mathrm{tot} }  
\! \left ( \boldsymbol{\zeta} \right ) $. 
Note that the recovered state in Eq.~\eqref{seq:gaus.meas.fb} is equivalent to what one would generate by a continuous weak measurement of $\hat x _{+} $, and then applying proper conditional feedback operations by the end of the measurement. However, we stress that our fully autonomous scheme does not require high-fidelity nonlocal measurement, making it compatible with near-term experimental platforms. Furthermore, our results uncover a previously unexplored class of dissipative dynamics with exotic entangling dynamics.

\subsubsection{State dynamics}

\label{suppsec:rec.dyn}

We now explicitly derive the dynamics of the mixed Gaussian state during the recovery protocol. To do this, let us first consider the mixed state after the autonomous measurement-and-feedforward operation, i.e., the final state 
$\hat \rho _{f} (t) $ after evolving according to the Lindbladian 
$\gamma  \mathcal{L} 
_{ \hat x _{+}  
\to \eta \hat y } $ (see Eq.~\eqref{eq:qme.meas.ff}) for time $t$, with 
\begin{align}
\label{seq:qme.mff.gen}
& \mathcal{L} 
_{ \hat x _{+}  
\to \eta \hat y  }
\hat \rho  
= - i 
[\eta \hat x _{+} 
\hat y , 
\hat{ \rho} ] 
+  
\mathcal{D}\left[ 
\hat x _{+} 
- i \eta \hat y \right ] 
\hat{ \rho}  
. 
\end{align}
In what follows, we assume initially all $3$ modes are in vacuum, i.e., we have, 
\begin{align}
\label{seq:rhof.ivac}
\hat \rho _{f} (t) 
= e ^{\gamma  \mathcal{L} 
_{ \hat x _{+}  
\to \eta \hat y  }
 t }
\left ( | \text{vac} \rangle
\langle \text{vac} | 
\right)
.  
\end{align}
However, we note that our results also generalize to other initial conditions. We omit the time variable $t$ in the following derivations when it does not cause confusion. We can now compute the characteristic function 
$\chi _{\rho _{f} } 
\! \left ( \boldsymbol{\xi  }
\right )$ of the Gaussian final state analytically, which is defined as  
\begin{align}
\label{seq:gaus.cf.def}
\chi _{\rho _{f} } 
\! \left ( \boldsymbol{\xi  }
\right )
\equiv 
\text{Tr} 
\left [  
\hat \rho _{f}  
\prod _{\ell =a , b, c} 
\hat D _{\ell }
\! \left ( \xi _{\ell } 
\right )
\right ] .
\end{align}
In Eq.~\eqref{seq:gaus.cf.def}, we introduce the standard displacement operators as 
\begin{align}
\hat{D} _{\ell }
\! \left ( \xi \right )
\equiv 
\exp ( \xi  
\hat  {\ell } ^{\dag}  
- \xi ^{*} 
\hat  {\ell } ) 
, \quad 
\ell =a, b, c 
. 
\end{align}
For a general Gaussian state, its characteristic function is also Gaussian, and the state can be fully specified via its first moments and the covariance matrix of quadrature operators. Let us first focus on the case with zero means (i.e., first moments of quadratures vanish), so that the characteristic function can be simply written in terms of the real, symmetric covariance matrix $\Sigma$ of the state, and the coefficient matrix $\Omega $ encoding the canonical commutation relations, as 
\begin{align}
&\chi _{\rho _{f} (t) } 
\left ( \boldsymbol{\xi } 
\right )
= \exp  \left ( 
- \frac{1}{2}
\boldsymbol{\xi } ^{T} 
\Omega \Sigma \Omega ^{T}
\boldsymbol{\xi } 
\right )
, \\
& \Sigma = \langle \{
\hat  r _{\sigma_{1} } 
,\hat  r _{\sigma_{2} } 
\} \rangle 
, \quad 
\hat  r _{\sigma  } 
\in \{ 
\hat x _{ \ell }
, \hat p _{ \ell }
\} ,
\\ 
& \Omega = \bigoplus 
_{ \ell =a, b, c } 
\omega , \quad 
\omega =\begin{pmatrix}
0 &1 \\
-1 & 0 
\end{pmatrix}
. 
\end{align}
Note that in this derivation, we redefine the mode $c$ quadratures as 
$(\hat y , \hat \pi) \to 
(\hat x _{c} , \hat p _{c}) $.
For the specific state in Eq.~\eqref{seq:rhof.ivac}, we can further explicitly express $\chi _{\rho _{f} (t) } 
\left ( \boldsymbol{\xi } 
\right )$ as 
\begin{align}
&\ln \left [
\chi _{\rho _{f} (t) } 
\left ( \boldsymbol{\xi } 
\right ) \right ]
\nonumber \\
= & - \frac{|\xi _{ a }|^2 
+ |\xi _{ b }|^2 
+ |\xi _{c}|^2
+ \gamma t  \left  
(\text{Re}\xi _{a}
+ \text{Re}\xi _{b} 
\right)^2 }{2} 
\nonumber \\
& - \gamma t \eta ^2
(1+ 2\gamma t )
(\text{Re}\xi _{c})^2 
\nonumber \\
& + 
\frac{  \gamma t
 \eta  }{\sqrt{2} }
(\text{Im}\xi _{a }
+ \text{Im}\xi _{b } )
\text{Re}\xi _{c}
\label{seq:FF_covmat_lnChi}
. 
\end{align}
For convenience, we also introduce the subsystem covariance matrices $\sigma _{ab}$ and 
$\sigma _{c}$ of modes $a$, $b$ and mode $c$, respectively, and the correlation matrix $\epsilon _{ab,c}$. The total covariance matrix can thus be expressed in terms of those submatrices as 
\begin{align}
\Sigma
=\begin{pmatrix}
\sigma _{ab} & \epsilon _{ab,c} \\
\epsilon _{ab,c} ^{T} & \sigma _{c} 
\end{pmatrix}
.  
\end{align}
The values of matrices $\Sigma$, or equivalently $\sigma _{ab}$, $\sigma _{c} $ and $\epsilon _{ab,c}$, can be directly read off from Eq.~\eqref{seq:FF_covmat_lnChi}.

We now consider the conditional $3$-mode state after the total system undergoes a POVM measurement described by Eq.~\eqref{seq:gaus.meas.povm}. As discussed, the conditional state corresponding to measurement result $\zeta$ is given by 
\begin{align}  
&\hat \rho 
_{\text{m}}
\! \left ( \zeta ; \mu \right ) 
= \frac{\hat{M}  _{\zeta, \mu} 
\hat \rho 
\hat{M}  _{\zeta; \mu} ^{\dag}}
{\text{Tr} \left (\hat{M}  _{\zeta, \mu} 
\hat \rho 
\hat{M}  _{\zeta, \mu} ^{\dag}  \right ) }
, 
\end{align}
where we define the POVM operators 
$\hat{M}  _{\zeta, \mu}$ as 
\begin{align}
\label{seq:gaus.meas3.op}
& \hat{M}  _{\zeta, \mu} 
= \frac{1 }{ \sqrt{2\pi} }
\hat{\mathbb{I} } _{ab} 
\otimes 
\hat{D}  _{c} 
\! \left ( \zeta  \right )
\hat{\Pi} _{c  } 
\! \left (  \mu  \right )
\hat{D} _{c} ^{\dag}
\! \left ( \zeta \right ) , \\
\label{seq:gaus.meas.ker}
& \hat{\Pi} _{c  } 
\! \left (  \mu  \right )
=  \hat{S}  _{c} 
\! \left (  \ln \mu  \right )
\left | 0 \rangle _{c} 
\langle 0 \right | 
\hat{S}  _{c} ^{\dag} 
\! \left (  \ln \mu  \right )
. 
\end{align}
Because $\hat{M}  _{\zeta, \mu}  $ is proportional to a projector, the resulting state 
$\hat \rho 
_{\text{m}}
\! \left ( \zeta ; \mu \right ) $ will be a product state between mode $a $, $ b$ and the corresponding pure state 
$ \hat{D}  _{c} 
\! \left ( \zeta  \right )
\hat{S}  _{c} 
\! \left (  \ln \mu  \right )
\left | 0 \right\rangle _{c}  
$ of mode $c$, as   
\begin{align}
& \hat \rho 
_{\text{m}}
\! \left ( \zeta ; \mu \right ) 
= \hat \rho _{ab} 
\! \left ( \zeta ; \mu \right ) 
\otimes 
\hat{D}  _{c} 
\! \left ( \zeta  \right )
\hat{\Pi} _{c  } 
\! \left (  \mu  \right )
\hat{D} _{c} ^{\dag}
\! \left ( \zeta \right ) 
. 
\end{align} 
We can thus rewrite the subsystem conditional state 
$\hat \rho _{ab} 
\! \left ( \zeta ; \mu \right ) $ of modes $a $, $b $ in terms of the total system conditional state 
$\hat \rho $, as 
\begin{align}
& \hat \rho _{ab} 
\! \left ( \zeta ; \mu \right ) 
= \text{Tr} _{c} 
\left ( \hat \rho 
_{\text{m}}
\! \left ( \zeta ; \mu \right ) 
 \right ) 
\\ 
= & \frac{
\text{Tr} _{c}  \left [
\hat{D}  _{c} 
\! \left ( \zeta  \right )
\hat{\Pi} _{c} 
\! \left (  \mu  \right )
\hat{D} _{c} ^{\dag}
\! \left ( \zeta \right ) 
\hat \rho 
\hat{D}  _{c} 
\! \left ( \zeta  \right )
\hat{\Pi} _{c} 
\! \left (  \mu  \right )
\hat{D} _{c} ^{\dag}
\! \left ( \zeta \right ) 
\right ] 
}
{ \text{Tr} \left [
\hat \rho 
\hat{D}  _{c} 
\! \left ( \zeta  \right )
\hat{\Pi} _{c  } 
\! \left (  \mu  \right )
\hat{D} _{c} ^{\dag}
\! \left ( \zeta \right )  
\right ] }
\nonumber \\ 
\label{seq:rho.cond12}
= & \frac{
\text{Tr} _{c} \left [
\hat \rho 
\hat{D}  _{c} 
\! \left ( \zeta  \right )
\hat{\Pi} _{c  } 
\! \left (  \mu  \right )
\hat{D} _{c} ^{\dag}
\! \left ( \zeta \right )  
\right ]
}{\text{Tr} \left [
\hat \rho 
\hat{D}  _{c} 
\! \left ( \zeta  \right )
\hat{\Pi} _{c} 
\! \left (  \mu  \right )
\hat{D} _{c} ^{\dag}
\! \left ( \zeta \right )  
\right ] }
. 
\end{align}

To show that Eq.~\eqref{seq:gaus.meas.fb} is correct, we now prove that the following equation holds for a generic $\zeta $ 
\begin{align}
\label{seq:rho.cond.disp}
& \hat \rho 
_{\text{m}}
\! \left ( \zeta ; \mu \right ) 
= \hat{D} _{\mathrm{tot} }  
\! \left ( \zeta \right )
\hat \rho _{\text{m}}
\! \left ( \zeta =0 ; 
\mu \right ) 
\hat{D}  _{\mathrm{tot} }   
^{\dag}
\! \left ( \zeta \right ) 
, \\
& \hat{D} _{\mathrm{tot} }  
\! \left ( \boldsymbol{\zeta} \right )
= \hat{D} _{a}  
\! \left ( \zeta _{a} \right )
\hat{D} _{b}  
\! \left ( \zeta _{b} \right )
\hat{D} _{c}  
\! \left ( \zeta \right ) 
, 
\end{align} 
where $\zeta _{a} $ and $\zeta _{b} $ are complex coefficients that can be determined using time evolution parameter $\gamma t $, the measurement projector $\hat{\Pi} _{c } 
\! \left (  \mu  \right )
$, and measurement result $\zeta $. To prove this, we first note that if we integrate out the first two modes from Eq.~\eqref{seq:rho.cond.disp}, the remaining equation for mode $3$ is automatically valid, since the POVM measurement operator in Eq.~\eqref{seq:gaus.meas3.op} is constructed via projectors $\hat{D}  _{c} 
\left ( \zeta  \right )
\hat{\Pi} _{c  } 
\! \left (  \mu  \right )
\hat{D} _{c} ^{\dag}
\left ( \zeta \right ) 
$. To show that Eq.~\eqref{seq:rho.cond.disp} holds for the subsystem state 
$\hat \rho _{ab} 
\! \left ( \zeta ; \mu \right )  $, we derive its characteristic function, which can be defined in analogy to Eq.~\eqref{seq:gaus.cf.def} as
\begin{align}
\label{seq:cf.cond12}
& \chi _{
\hat \rho _{ab} 
\! \left ( \zeta ; \mu \right ) } 
\! \left ( 
\boldsymbol{\xi } _{ab} 
\right )
\nonumber \\
\equiv &
\text{Tr} 
\left [  
\hat \rho _{ab} 
\! \left ( \zeta ; \mu \right ) 
\hat D _{ a }
\! \left ( \xi _{a } 
\right )
\hat D _{b}
\! \left ( \xi _{b} 
\right )
\right ] .
\end{align}
Note that we use subscript in 
$\boldsymbol{\xi } _{ab} $ to denote explicitly that LHS of Eq.~\eqref{seq:cf.cond12} is characteristic function of the $2$-mode ($a$, $b$) subsystem state. Making use of Eq.~\eqref{seq:rho.cond12}, we obtain 
\begin{align} 
& \chi _{
\hat \rho _{ab} 
\! \left ( \zeta ; \mu \right ) } 
\! \left ( \boldsymbol{\xi }
_{ab} 
\right )
\nonumber \\
= & \frac{ 
\text{Tr} _{c} \left [
\hat \rho 
\hat D _{ a }
\! \left ( \xi _{a } 
\right )
\hat D _{b}
\! \left ( \xi _{b } 
\right )
\hat{D}  _{c} 
\! \left ( \zeta  \right )
\hat{\Pi} _{c  } 
\! \left (  \mu  \right )
\hat{D} _{c} ^{\dag}
\! \left ( \zeta \right )  
\right ]
}{\text{Tr} \left [
\hat \rho 
\hat{D}  _{c} 
\! \left ( \zeta  \right )
\hat{\Pi} _{ c } 
\! \left (  \mu  \right )
\hat{D} _{c} ^{\dag}
\! \left ( \zeta \right )  
\right ] }
 .
\end{align}
To proceed, we denote the correlation matrix of measurement kernel $\hat{\Pi} _{c} 
\! \left (  \mu  \right ) $ as 
$ V _{ \mu }  $. One can thus show that the subsystem characteristic function takes the following form 
\begin{align} 
& \chi _{
\hat \rho _{ab} 
\! \left ( \zeta ; \mu \right ) } 
\! \left ( 
\boldsymbol{\xi } _{ab} 
\right )
\nonumber \\ 
= & \exp  \left \{ 
- \frac{1}{2}
\boldsymbol{\xi } ^{T} _{ab} 
\Omega  _{ab}  
\tilde{\sigma} _{ab} \Omega _{ab}^{T}
\boldsymbol{\xi } _{ab} 
- i \overline{r} _{ab} 
 ^{T}
\Omega _{ab}^{T}
\boldsymbol{\xi } _{ab}
\right \}
,  
\end{align}
where the new conditional state covariance matrix $\tilde{\sigma} _{a} $ and first moment averages are given by 
\begin{align}
\label{seq:covmat.cond12}
& \tilde{\sigma} _{ab} 
= \sigma _{ab} - \epsilon _{ab,c}  
\left ( 
\sigma _{c}  + 
V _{ \mu } 
\right ) ^{-1} 
\epsilon _{ab,c} ^{T}  
, \\
\label{seq:mean.cond12}
& \overline{r} _{ab}  ^{T}
= - \zeta^{T} 
\left ( 
\sigma _{c} 
+ V _{ \mu } 
\right ) ^{-1} 
\epsilon _{ab,c} ^{T}  
. 
\end{align}
Note that the covariance matrix is deterministic and does not depend on value of the measurement result $\zeta$. Thus, we have proved Eq.~\eqref{seq:rho.cond.disp}, where the displacement amplitudes 
$\zeta _{a} $, $\zeta _{b} $ can be extracted from Eq.~\eqref{seq:mean.cond12}. For the specific case where 
$ \hat{\Pi} _{c } 
\! \left (  \mu  \right )  $ corresponds to a pure state squeezed in quadrature $\hat \pi$, 
$ V _{ \mu }  $ simplifies to a diagonal matrix, which can be written as 
\begin{align}
V _{ \mu } 
= \text{diag} (\mu ^{-1}, \mu)
. 
\end{align}
In this case, the displacement amplitudes are given by 
\begin{align}
& \zeta _{a}= 
\zeta _{b}= 
-\frac{ \gamma t \eta 
\sqrt{2 } }
{\mu 
+ 1+ 2 \gamma t \eta ^2
(1+ 2 \gamma t ) } 
\text{Im} \zeta 
. 
\end{align}

\subsection{Conditional entanglement generation due to collective measurement and a single-site detuning Hamiltonian}

\label{suppsec:meas.H.cond}

As discussed in the main text, the interplay between a nonlocal continuous measurement on $\hat {x} _{+}$ and a local detuning Hamiltonian could result in two distinct asymptotic regimes of entanglement growth. Here, we provide physical intuitions for understanding such behavior. For convenience, we again express the local Hamiltonian as being perturbed from the quantum-mechanics-free subsystem (QMFS) regime (see Eq.~\eqref{eq:h.qmfs.perturb})
\begin{align}
\label{seq:h.qmfs.perturb}
&  \hat H _{\mathrm{det}}
= (\omega + \delta \omega ) \hat a ^{\dag}
\hat a 
+ ( - \omega + \delta \omega ) \hat b ^{\dag}
\hat b
. 
\end{align}
If $\delta \omega \ne 0$, the long-time entanglement generation generally saturates. This can be intuitively understood as in each trajectory, the nonlocal measurement effectively squeezes the collective quadrature $\hat {x} _{+}$, whereas the Hamiltonian couples it to other non-commuting quadratures, which would interfere with the squeezing dynamics and cut off long-time entanglement growth. However, if 
$\delta \omega = 0 $, the conditional state undergoes logarithmic entanglement growth. Specifically in the QMFS regime ($\delta \omega = 0, \omega \ne 0 $),  we obtain unbounded entanglement generation, with the long-time asymptotic entanglement grows twice as fast as the Hamiltonian-free case (see Eq.~\eqref{eq:meas.cond.xp.ent}), as shown in Fig.~\ref{fig:qmfs.det}(a). In this case, it is useful to rewrite the QMFS Hamiltonian as~\cite{Caves2012,Polzik2017}
\begin{align}
\label{seq:h.qmfs}
\hat H _{\mathrm{QMFS}}
= \frac{\omega }{2} 
( \hat x _{+} \hat x _{-} 
+ \hat p _{+} \hat p _{-} )
,
\end{align}
where we define the collective quadratures as 
$\hat x _{\pm} \equiv 
(\hat x _{a} \pm \hat x _{b})/\sqrt{2}$ and $\hat p _{\pm} \equiv 
(\hat p _{a} \pm \hat p _{b})/\sqrt{2}$. 
Thus, enhancement in entanglement generation can be explained using the QMFS structure that couples $\hat {x} _{+}$ to a commuting quadrature $\hat {p} _{-}$, so that in the long-time limit, the measurement process effectively squeezes both quadratures.

Let us now consider the opposite limit, where the long-time conditional entanglement asymptotically vanishes with a one-sided detuning  
$\delta \omega = \pm \omega$ (see, e.g., the green curve in Fig.~\ref{fig:qmfs.det}(a)). To obtain a physical understanding of the suppression of conditional entanglement in this finely tuned limit, we note that both the Hamiltonian in Eq.~\eqref{seq:h.qmfs.perturb} and measured operator $\hat {x} _{+}$ now commute with the local quadrature $\hat {x} _{b}$. As such, the conditional dynamics can be viewed as due to the combination of an effective measurement-induced squeezing in $\hat {x} _{b}$ (with the covariance 
$\langle \delta \hat {x} _{b} ^{2}\rangle 
\propto t ^{-1} $ in the $ t \to \infty $ limit) and other oscillatory dynamics. In the long-time limit, the local measurement-induced squeezing effect dominates, so that the net effect of measurement of $\hat {x} _{+}$ and a single-mode detuning Hamiltonian leads to zero long-time entanglement.

\subsection{Simulation results for the adaptive protocol with different number of registers}

\label{suppsec:qmfs.sliced}

\begin{figure}[t]
    \centering
    \includegraphics[width=\columnwidth]{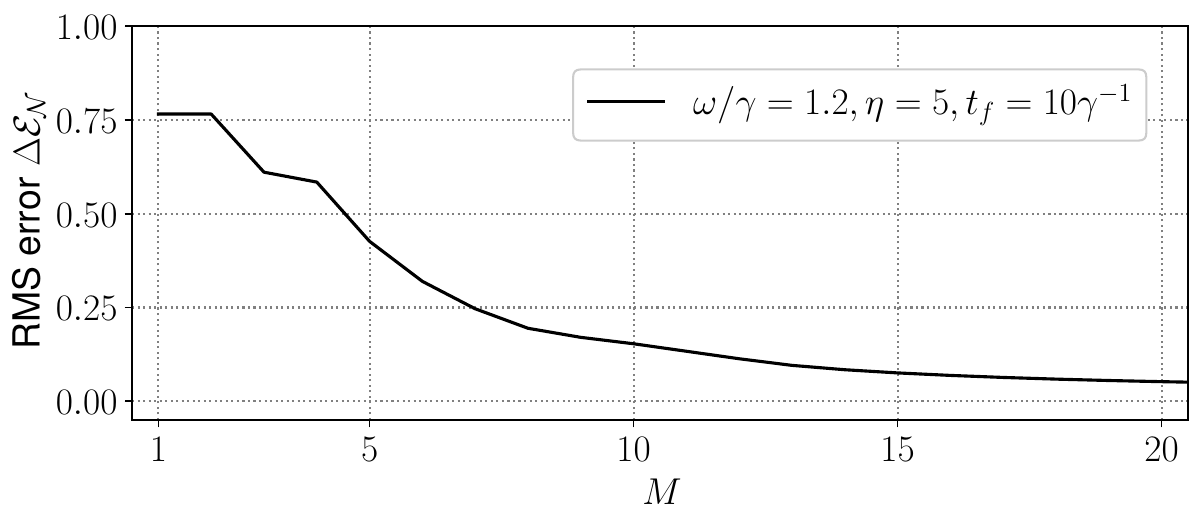}
    \caption{Root-mean-squared (RMS) error in entanglement negativity $\Delta \mathcal{E} _{\mathcal{N}}$ (see Eq.~\eqref{seq:qmfs.rms.err}) between the postselected entanglement generated by continuous monitoring of $\hat {x} _{+} $ and Hamiltonian in Eq.~\eqref{seq:h.qmfs.perturb}, versus the deterministic entanglement generated by the corresponding measurement-free deterministic dynamics (c.f.~Eq.~\eqref{eq:qme.meas.ff.nmodes} in the main text) for different number of registers $M$ averaged over range of detunings $\delta \omega / \gamma \in [0,1.5]$ (as shown in Fig.~\ref{fig:qmfs.det}(b)), both evaluated at a fixed evolution time $t_{f}$.  
    Parameters: $\omega / \gamma = 1.2$, $\eta =5 $, $t _{f}=10 \gamma^{-1}$. 
    }
    \label{suppfig:qmfs.det.rmserr}
\end{figure}

In the main text, we show that the adaptive protocol in  Eq.~\eqref{eq:qme.meas.ff.nmodes} can capture nontrivial features in the measurement-induced conditional entanglement due to the interplay between a nonlocal continuous measurement on $\hat {x} _{+}$ and a local Hamiltonian perturbed from the QMFS regime (see Eq.~\eqref{seq:h.qmfs.perturb}).  
As discussed, to achieve this, we need multiple auxiliary register modes $ c _{\ell}$ 
($\ell =1, 2,\ldots, M$), so that the feedforward protocol in Eq.~\eqref{eq:qme.meas.ff.nmodes} involves sequentially coupling the system modes $a,b$ to each of the register modes. Here we provide more details on the root-mean-squared error analysis, and the effect of using smaller number of registers.

Moreover, the deviation between the adaptive-dynamics-generated entanglement and the postselected entanglement decreases as we increase $M$, and this error becomes negligible for sufficiently large $M$. 
Indeed, as shown in Fig.~\ref{suppfig:qmfs.det.rmserr}, the root-mean-squared error 
\begin{align}
\label{seq:qmfs.rms.err}
\Delta \mathcal{E} _{\mathcal{N}}
\equiv \sqrt{ 
\overline{ (\mathcal{E} _{\mathcal{N}}^{(\mathrm{ps})} (\delta \omega)
- \mathcal{E} _{\mathcal{N}}^{(\mathrm{deter})}
(\delta \omega))^2 } }
,
\end{align}
between the postselected entanglement, $\mathcal{E} _{\mathcal{N}}^{(\mathrm{ps})}$, and the unconditional entanglement in the adaptive protocol, $\mathcal{E} _{\mathcal{N}}^{(\mathrm{deter})}$, decreases rapidly as we increase $M$, and $M=20$ provides a suitable choice for capturing conditioned entanglement in this case. The overline symbol in Eq.~\eqref{seq:qmfs.rms.err} denotes an average over different values of the system Hamiltonian parameter $\delta \omega / \gamma$ used in the simulations. However, we note that for all examples considered here and in Sec.~\ref{appsec:multi.ff.params}, the specific value of $\delta \omega / \gamma$ does not change the qualitative trend of dependence of $\Delta \mathcal{E} _{\mathcal{N}}$ on $M$ (or feedforward parameter $\eta$).

\begin{figure}[t]
    \centering
    \includegraphics[width=\columnwidth]{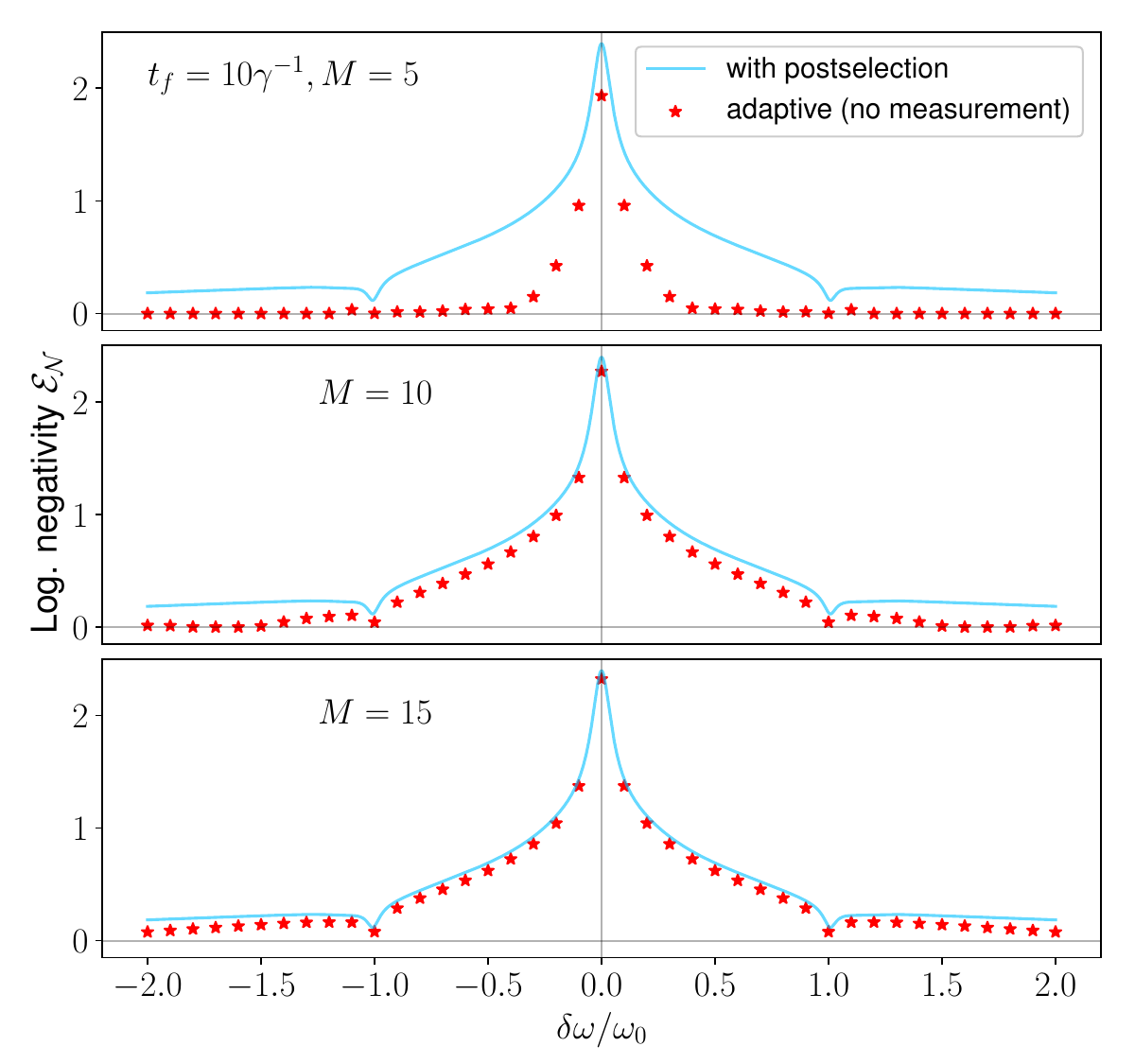}
    \caption{Conditional entanglement (light blue curves) due to simultaneous measurement of $\hat {x} _{+} $ and local Hamiltonians given by Eq.~\eqref{eq:h.qmfs.perturb} and deterministic entanglement generation due to the adaptive protocol (red asterisks, see Eq.~\eqref{eq:qme.meas.ff.nmodes}), when using varying numbers of registers, $ M =5,10 $ or $15$. The bipartition is taken between mode $ a$ and the rest of the system. The deterministic dynamics mimic a process where we divide the total evolution time $t _{f} =10 \gamma^{-1} $ into $M$ equal segments, and then sequentially perform measurement-and-feedforward dynamics from modes $a $ and $ b $ to one of the register modes. Other parameters: $\omega / \gamma = 1.2$, $\eta =5$. 
    }
    \label{suppfig:qmfs.det.regdim}
\end{figure}

Here we provide additional simulation results on the adaptive protocol with smaller values of $M$. That is, we compute the entanglement of the postselected dynamics due to simultaneous weak continuous $\hat {x} _{+}$ measurement and Hamiltonian 
$\hat H _{\mathrm{det}}$ evolution, and we compare it with the deterministic entanglement due to the feedforward protocol in Eq.~\eqref{eq:qme.meas.ff.nmodes} with different $M$. As shown in Fig.~\ref{suppfig:qmfs.det.regdim}, the deterministic state already captures the transition at the QMFS 
($\delta \omega=0$) point and the vanishing of entanglement at special tuning point $\delta \omega=\omega$ with even a smaller number of registers such as $M=10$.

For capturing entanglement due to more generic measurement dynamics (with postselection), we estimate the number of $M$ required to be given by the product of total evolution time and the typical frequency scale; for the dynamics considered here, we thus expect that $\gamma t_f =10$ register modes are necessary, which is consistent with numerical simulations. For the general purpose of detecting measurement-induced entanglement transitions (MIPT), we conjecture that the number of register modes required can scale sublinearly with the system size, which would enable a scalable test of MIPT.

\subsection{Generalization to bosonic lattice systems}

\label{suppsec:multimode}

In the main text, we discussed the generalization of our measurement-free adaptive protocol to many-body systems. More specifically, we consider a multimode setup consisting of a $1$D bosonic lattice with additional auxiliary registers attached to each measurement bond, and show that the deterministic entanglement generated by the measurement-free dynamics captures spatial structure of conditional entanglement growth due to the target measurement-induced process. Here, we provide a more detailed discussion on features of the conditional entanglement generation, as well as the specific parameters used to construct the measurement-free protocol.

\begin{figure}[t]
    \centering
    \includegraphics[width=\columnwidth]{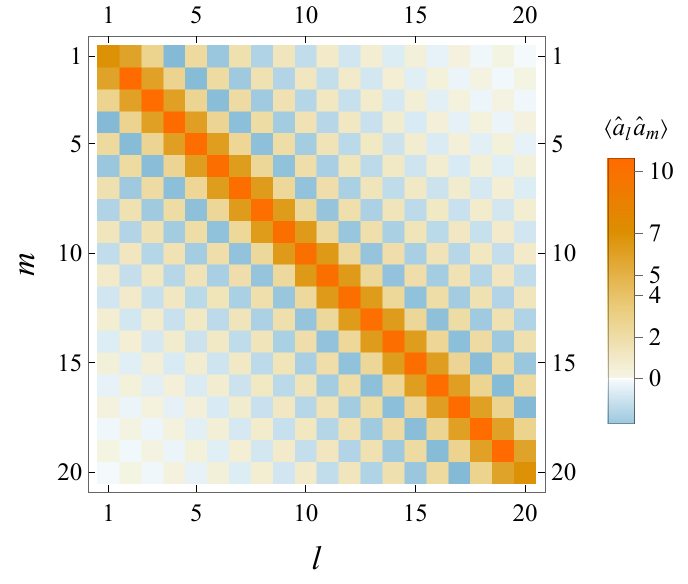}
    \caption{Pairing correlators 
    $ \langle \hat a _{l} \hat a _{m} \rangle$ of the multimode squeezed state generated by the conditioned dynamics due to simultaneously measuring the bond quadratures $\hat m _{j}  
= (\hat x _{j } 
+  \hat x _{j +1 })/\sqrt{2} $ on a $n$-mode bosonic lattice. Here we take lattice size $n=20$ and evolution time $t _{f}=10 \gamma^{-1}$.  
    }
    \label{suppfig:cond.qnd.multi}
\end{figure}

\subsubsection{Hamiltonian-free dynamics: logarithmic-in-time entanglement growth and unbounded increase of the correlation length}
\label{appsec:multi.meas.only}

For simplicity, we first consider the Hamiltonian-free measurement-only regime of the lattice model discussed in the main text. More concretely, we consider $n$ bosonic modes $a _{j}$ arranged in a periodic 1D lattice as the target system, undergoing continuous monitoring of operators 
$\hat m _{j}  
= (\hat x _{j } 
+  \hat x _{j +1 })/\sqrt{2} $
on each nearest neighbor bond. For convenience, we assume $n$ is even throughout the discussion.
We assume a vacuum initial state for concreteness, so that we can again analytically solve the exact postselected state dynamics. By diagonalizing the coefficient matrix associated with the quadratic sum $ \sum _{j} \hat m _{j} ^{2} $ via an orthogonal matrix $[o _{jk} ]$ and a set of $n$ positive coefficients $\lambda  _{j} $, we have
\begin{align}
\label{seq:qnd.multi.spec.quad}
& \sum _{j=1} ^{n} 
\hat m _{j}  ^{2} 
= \sum _{j=1} ^{n} 
\lambda  _{j} 
\hat {\tilde{x}} _{j}^{2} 
, \quad 
\hat {\tilde{x}} _{j}
=
\sum _{k=1} ^{n} 
o _{jk} 
\hat x _{k} 
,   
\end{align}
and the conditional state in the asymptotic long-time regime can be written as
\begin{align}
\label{seq:Psi.nmode.multi.tb}
\gamma t \gg 1: \quad 
\left | \Psi _{n} 
\right \rangle
& = e 
^{ \frac{ r  }{ 2 } 
\sum _{j=1} ^{n }   
\left ( \hat d _{j}  ^{2}
-\hat d _{j}  ^{\dag 2} 
\right) }  
\left | 0 \right \rangle
,     
\end{align}
where the effective squeezing mode operators are given by $\hat d _{j} 
= \sum _{j=1} ^{n} 
o _{jk} 
\hat a _{k} $, and the squeezing parameters 
$r \sim \log( \gamma t )$. Note that Eq.~\eqref{eq:meas.cond.xp.ent} can be viewed as a single-mode special case of Eq.~\eqref{seq:Psi.nmode.multi.tb}. For a larger lattice size, we generally expect long-range entanglement correlations. See Fig.~\ref{suppfig:cond.qnd.multi} for the pairing correlators 
$\langle \hat a _{l}\hat a _{m} \rangle$ in a $1$-dimensional chain of bosonic modes (with $n=20$ modes in the lattice) undergoing measurements of $\hat m _{j} $, which exhibits long-range correlations; in this case, the pairing correlators are equal to the corresponding normal-ordered correlators 
$\langle \hat a _{l}\hat a _{m} \rangle
= \langle \hat a ^{\dag}_{l}\hat a _{m} \rangle $. States of the form in Eq.~\eqref{seq:Psi.nmode.multi.tb} are generally known as H-graph states~\cite{vanLoock2011}, and can be used to generate continuous variable cluster states.

\begin{figure}[t]
    \centering
    \includegraphics[width=\columnwidth]{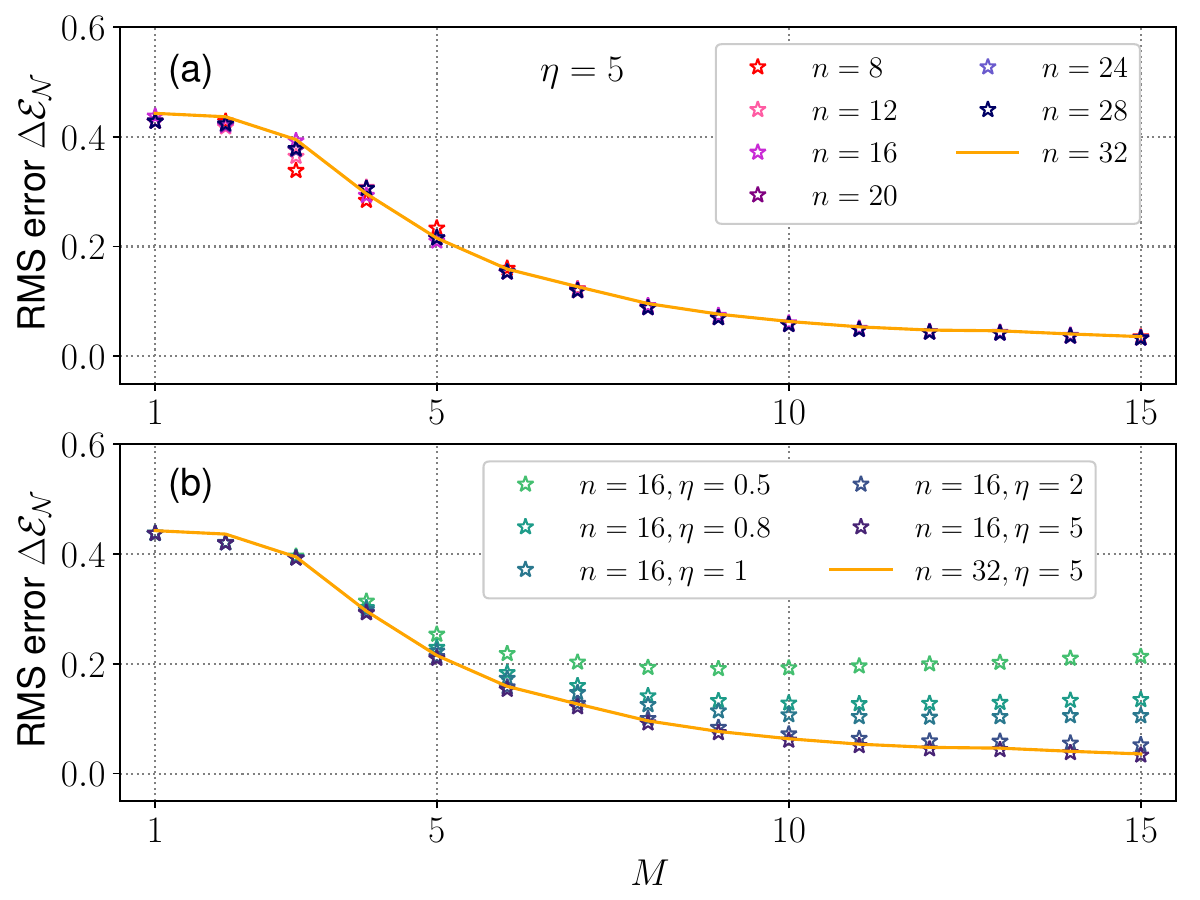}
    \caption{RMS error in entanglement negativity $\Delta \mathcal{E} _{\mathcal{N}}$ (see Eq.~\eqref{seq:qmfs.rms.err}) between the postselected entanglement generated by continuous monitoring of $\hat m _{j}  $ and Hamiltonian in Eq.~\eqref{seq:h.qmfs.perturb.multi}, versus the deterministic entanglement generated by the corresponding measurement-free deterministic dynamics (c.f.~Eq.~\eqref{seq:qme.meas.ff.nmodes}, see also Eq.~\eqref{eq:qme.meas.ff.nmodes} in the main text) for varying number of registers $M$ per bond. Here the entanglement negativities are evalutated with respect to the half-system bipartition. All data points are taken to be averaged over the range of detunings $\delta \omega / \gamma \in [0.1,0.5]$, and for each value of $\delta \omega$, we choose the evolution time $t _{f}$ such that the deterministic entanglement generation by the measurement-free protocol has stabilized. Other parameters: (a)  $\eta =5 $ with various total system sizes $n=8,12,16,20,24,28,32$; (b) $\eta =0.5,0.8,1,2,5 $ for $n=16$, which are compared with the baseline value at $\eta = 5 $ and $n=32$.
    In both panels, we take $\gamma t _{f}= (40,26,21,16,15) $ for $\delta \omega / \gamma = (0.1,0.2,0.3,0.4,0.5)$. 
    }
    \label{suppfig:nreg.multi}
\end{figure}

We now consider the corresponding measurement-free dissipative protocol. As the target conditional dynamics only consist of commuting measurement on each bond in the lattice, we introduce $n$ auxiliary modes (one for each bond) as the register.
We focus on a setup where on each bond of the lattice, we apply a dissipative Lindbladian of the form Eq.~\eqref{eq:qme.meas.ff}, which mimics measuring the quadrature on each bond 
$\hat m _{j}  
= (\hat x _{j } 
+  \hat x _{j +1 })/\sqrt{2} $ and then applying conditional feedforward forces to the corresponding register quadrature 
$ \hat y _{j} $. The system dynamics is thus given by the following master equation
\begin{align}
\label{seq:qme.nmode.tb}
& \frac{d\hat \rho  }{d t}
= \gamma \sum _{j=1} ^{n} 
\mathcal{L} 
_{ \hat m _{j}  
\to \eta \hat y _{j}  }
\hat \rho 
, \quad
\hat m _{j}  
= \frac{\hat x _{j } 
+  \hat x _{j +1 } }{\sqrt{2}}
. 
\end{align}
The Lindbladian is defined similarly as Eq.~\eqref{eq:qme.meas.ff} as 
$\mathcal{L} 
_{ \hat m _{j}  
\to \eta \hat y _{j}  }
= - i 
[\eta \hat m _{j}  
\hat y _{j} , 
\hat{ \rho} ] 
+  
\mathcal{D}\left[ 
\hat m _{j}  
- i \eta \hat y _{j} \right ] 
\hat{ \rho}  
$.

In direct analogy to the two-mode example discussed in the main text (with no Hamiltonian), 
the dynamics in Eq.~\eqref{seq:qme.nmode.tb} can deterministically reproduce the entanglement one would obtain from measuring the quadratures $ \hat m _{j}  $ simultaneously and then postselecting to a particular set of measurement records. The resulting final state from this adaptive protocol is mixed, but we can apply a recovery operation similarly as that discussed in Eqs.~(\ref{seq:gaus.meas.povm}-\ref{seq:gaus.meas.fb}) to recover an approximately pure, multimode-entangled state. The recovery now involves measuring each of the register mode quadratures $ \hat \pi _{j}  $ (defined via the canonical commutation relations 
$ [ \hat y _{j} , \hat \pi _{k} ]
= i \delta _{jk} $) and applying conditional local feedback forces to the corresponding target mode quadratures 
($ \hat x _{j}  $ and $ \hat x _{j+1}  $).

\subsubsection{Measurement-free protocol with onsite Hamiltonian dynamics: dependence of performance on the number of registers}

\label{appsec:multi.ff.params}

The dynamics considered in Sec.~\ref{appsec:multi.meas.only} only involves continuous measurements of commuting quadrature operators
$\hat m _{j}  
= (\hat x _{j } 
+  \hat x _{j +1 })/\sqrt{2} $, and generates logarithmic-in-time entanglement growth. We now introduce an additional on-site Hamiltonian that can interfere nontrivially with the measurement-induced entanglement growth, as  (assuming $n$ is even; see also Eq.~\eqref{seq:h.qmfs.perturb})
\begin{align}
\label{seq:h.qmfs.perturb.multi}
& \hat H _{\mathrm{det,multi}} = \sum _{j=1} ^{ {n}/{2}}  \hat H _{j}, 
\\
\hat H _{j} = &  (\omega + \delta \omega ) \hat a _{2j}^{\dag}
\hat a _{2j} + ( \delta \omega - \omega ) \hat a _{2j-1} ^{\dag}
\hat a _{2j-1} 
. 
\end{align}
To illustrate the performance of the measurement-free protocol, we focus on the regime with uniform detuning, i.e.~$\delta \omega \ne 0, \omega =0$, but we note that our approach is also applicable to the more general cases.  
As discussed in the main text, we can directly generalize the measurement-free protocol in Eq.~\eqref{eq:qme.meas.ff.nmodes} to the multimode regime. More specifically, we consider the dissipative dynamics involving $n$ system modes, with $M$ registers per measurement bond, as 
\begin{align}
\label{seq:qme.meas.ff.nmodes}
 (d \hat \rho  / dt) = & 
- i [ \hat H _{\mathrm{det,multi}} 
, \hat \rho ]
\nonumber \\
& + \gamma \sum _{j =1} ^{ n } \sum _{\ell =1} ^{ M } 
f_{\ell} (t; t_{f}) \mathcal{L} 
_{ \hat m _{j}  
\to \eta \hat y _{j; \ell} }  
\hat \rho 
, 
\end{align}
where $f_{\ell} (t; t_{f}) 
\equiv 
\Theta (t-\frac{\ell-1}{M} t_{f}) 
\Theta (\frac{\ell}{M} t_{f} - t ) $ 
denotes the coarse-graining function that enables a directional coupling from system operator $\hat m _{j}  $ at $j$th measurement bond to the register mode $\hat{c}_{j,\ell}$ during the $\ell$th time interval.

To investigate the effect of different number of registers, we compute the root-mean-squared (RMS) error (see Eq.~\eqref{seq:qmfs.rms.err}) between the postselected entanglement and the unconditional entanglement in the adaptive protocol numerically; the results are shown in Fig.~\ref{suppfig:nreg.multi} for varying number of system lattice size $n$, the number of registers per bond $M$, and the feedforward parameter $\eta$. 
We note that for the measurement-free protocol to effectively capture the postselected entanglement generation, the required value of $\eta $ and number of register modes $M$ per measurement bond do not increase with system size $n$. Our protocol thus introduces a linear-in-system-size hardware overhead due to the use of the auxiliary modes.

\subsection{Alternative examples of directional adaptive dynamics in multiqubit(qudit) systems}

\label{suppsec:mff.dv}

Here, we provide more examples of our directional adaptive dynamics based on discrete variable systems (i.e., qudits), which can be viewed as direct generalizations of Eq.~\eqref{eq:qme.mff.2qb}.

\begin{figure}[t]
    \centering
    \includegraphics[width=\columnwidth]{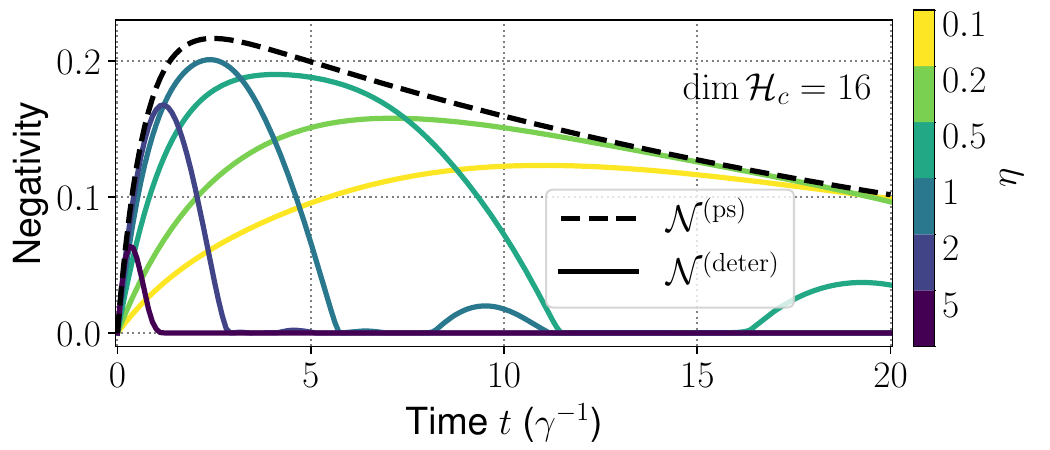}
    \caption{Deterministic entanglement generation by Eq.~\eqref{eq:qme.mff.2qb}, between qubit $1$ and the rest of the system-register, via the adaptive process with different $\eta$ (solid curves). Note that in the short-time regime, the deterministic entanglement monotonically approaches the conditional case (dashed curve) with increasing $\eta$. 
    }
    \label{fig:qudit_2023Jan25}
\end{figure}

We start with a discussion on the effect of varying the feedforward $\eta$ parameter. For this, we focus on the setup given by Eq.~\eqref{eq:qme.mff.2qb} in the main text, and compute the deterministic entanglement growth for different $\eta$. As shown in Fig.~\ref{fig:qudit_2023Jan25}, in the very-short-time regime, the effect of $\eta$ for qudit registers is similar to the case with a harmonic oscillator register: the deterministic entanglement monotonically increases as $\eta$ increases. However, for the qubit-qudit dynamics in Eq.~\eqref{eq:qme.mff.2qb}, the benefit from using a large $\eta$ can quickly die off at longer times, which is due to the finite size effect of the register. For a generic adaptive protocol (of the form Eq.~\eqref{eq:qme.mff.2qb}) with a finite dimensional register, we expect a trade-off between those two effects. Regardless, if we choose an $\eta$ of $o(1)$, the intermediate- and long-time entanglement would not be limited by the value of $\eta$ (and will instead be limited by, e.g., the dimension of register). This is the regime in which we choose to plot  Fig.~\ref{fig:qubit_mff_trunc_ho}.

\begin{figure}[t]
    \centering
    \includegraphics[width=\columnwidth]{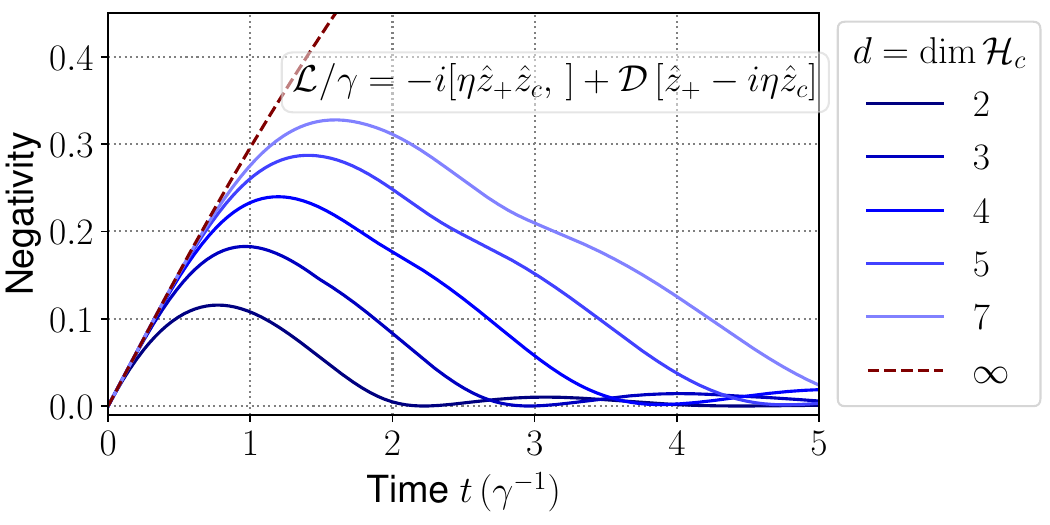}
    \caption{Entanglement generated between qudits $1$ and $(2,3)$ due to measurement-and-feedforward process for discrete variable systems. The total system consists of three qudits with Hilbert space dimension $d ={\mathrm{dim}} \mathcal{H} _{c} $ ($={\mathrm{dim}} \mathcal{H} _{a} ={\mathrm{dim}} \mathcal{H} _{b}$) ranging from $2$ to $7$ (see legend), whose Lindbladian is given by Eq.~\eqref{eq:qme.meas.ff} with the oscillator Hilbert space truncated up to $ d $-photon Fock state (see Eq.~\eqref{seq:qme.mff.qb} for the qubit case); the initial state corresponds to the product vacuum state. Other parameter: $\eta =1 $.  }
    \label{fig:qd_ent}
\end{figure}

Let us now consider a slightly different setup, consisting of a target system with $2$ qubits and a single-qubit register, where the first two qubits correspond to subsystems $A$ and $B$, respectively. Inspired by Eq.~\eqref{eq:qme.meas.ff}, one can write down a qubit version of the autonomous measurement-and-feedforward process, simply by replacing the position quadrature operators 
$\hat x _{\ell} $ with qubit Pauli operators 
$\hat \sigma _{x, \ell} / \sqrt{2}  $. Thus, the system dynamics can be desribed by master equation 
$(d \hat \rho / d t ) = \mathcal{L} \hat \rho$, with the Lindbladian 
\begin{align}
\label{seq:qme.mff.qb}
\mathcal{L} \hat \rho
= & - i \gamma
[\eta \hat z _{+} 
\hat z _{c} , 
\hat{ \rho} ] 
+ \gamma 
\mathcal{D}\left[ 
\hat z _{+} 
- i \eta \hat z _{c} \right ] 
\hat{ \rho}   
, 
\end{align}
with $\hat z _{+} 
= (\hat \sigma _{x, a }  
+ \hat \sigma _{x, b }  )/2$, and 
$\hat z _{c} = \hat \sigma _{x, c } /\sqrt{2} $ denoting the measurement and feedforward operators, respectively, and $\eta $ quantifying the feedforward strength.
For reasons that will become clear later, we focus on the case with the initial state given by 
$\left |\downarrow \downarrow \downarrow  \right\rangle $, where 
$ \left |\downarrow  \right\rangle $ is the ground state of Pauli $z$ operator, 
$\hat \sigma _{z } = 
\left |\uparrow \rangle 
\langle \uparrow \right |
- \left |\downarrow \rangle 
\langle \downarrow \right | $. It is then straightforward to see that the action of bosonic quadrature operator $\hat x _{\ell} $  on vacuum state can be directly mapped to how the Pauli $x$ operator acts on qubit ground state 
$\left |\downarrow  \right\rangle $ up to a scaling factor, as (omitting mode and qubit indices for notational simplicity)
\begin{align}
\hat x  
\left | 0 \right\rangle 
= \frac{1}{\sqrt{2} }
\left | 1 \right\rangle 
\leftrightarrow
\frac{\hat \sigma _{x } }{\sqrt{2} }
\left |\downarrow  \right\rangle 
= \frac{1 }
{\sqrt{2} }
\left | \uparrow \right\rangle 
. 
\end{align}
Hence, in the short-time limit 
$\gamma t \ll 1 $, the dynamics of the oscillator system provide a good approximation for the qubit dynamics. We can further construct higher dimensional versions of the adaptive dynamics Eq.~\eqref{seq:qme.mff.qb}, by replacing all the qubit operators in Eq.~\eqref{seq:qme.mff.qb} with truncated oscillator quadrature operators (to dimension $d$). In Fig.~\ref{fig:qd_ent}, we plot the entanglement negativity in qubit system with respect to the bipartition between the first and the other two qubits (solid darkest blue curve), the case with truncated oscillators and a similar bipartition (other light blue curves), as well as the corresponding linear bosonic dynamics (dashed red). Note that all curves overlap near $t=0 $. We expect the adaptive protocol with a finite-dimensional register to generally be efficient in this short-time and large-$\eta$ regime (see also Fig.~\ref{fig:qudit_2023Jan25}).

Note that, similar to the finite-dimensional system case in the main text, this correspondence will break down as evolution time increases, and the qudit entanglement generated by the adaptive protocol will eventually deviate from the postselected state entanglement at longer times. This can be understood as due to the fact that the register does not have an infinite phase space. For example, for the qubit case in Eq.~\eqref{seq:qme.mff.qb}, a convenient physical picture to use here is given by the register Bloch sphere. More concretely, as the qudit evolves in time, nonlinear effects will become important, causing the third qubit's trajectory state wavefunction to wrap around in Bloch sphere, which leads to effective loss of information about measurement result and hence degradation of entanglement in the averaged state over all trajectories. The timescale of the maximum qubit entanglement negativity can be connected to the time after which the qubit wavefunction starts to experience curvature of the Bloch sphere. As shown in Fig.~\ref{fig:qd_ent}, the maximal entanglement time monotonically increases with growing Hilbert space dimension of the constituent qudits.

\subsection{Details on physical implementation of the measurement-free protocol}

\label{suppsec:phys.impl}

Here we discuss the necessary elements required for the experimental implementation of the adaptive protocol discussed in the main text. The key setup is given by Eq.~\eqref{eq:qme.meas.ff}, which describes an autonomous measurement-and-feedforward process in a $3$-mode system, as 
\begin{align}
\label{seq:qme.meas.ff}
&  \mathcal{L} 
_{ \hat x _{+}  
\to \eta \hat y  }
\hat \rho  
\equiv 
- i 
[\eta \hat x _{+} 
\hat y , 
\hat{ \rho} ] 
+  
\mathcal{D}\left[ 
\hat x _{+} 
- i \eta \hat y \right ] 
\hat{ \rho}  
. 
\end{align}
Depending on properties of the physical system, there are two possible strategies to achieve this type of dynamics. First, for physical systems that inherently allow nonreciprocal QND interactions, e.g.,~in levitated nanoparticles~\cite{Rieser2022}, one can directly engineer the desired dissipative and Hamiltonian interactions. Second, and more generally, for systems that do not naturally preserve the QND structure, one can adopt a synthetic approach and separately engineer the coherent and dissipative couplings. This reservoir-engineering approach has been demonstrated in a range of experimental platforms, including e.g.~superconducting circuits, atomic ensembles~\cite{Polzik2011}, trapped ions~\cite{Zoller1996,Home2015,Home2019}, coupled photonic or phononic oscillators, and optomechanical systems~\cite{Schwab2015,Sillanpaa2015,Schwab2016}. For each physical device, the first route realizes a specific version of Eq.~\eqref{seq:qme.meas.ff}, whereas the second approach in principle can be used to engineer a tunable class of quantum master equations with the desirable form.

We now provide a detailed discussion on the synthetic approach. The basic idea is to implement the collective dissipator in Eq.~\eqref{seq:qme.meas.ff} via an auxiliary reservoir mode $z$; the Hamiltonian term in Eq.~\eqref{seq:qme.meas.ff} can also be engineered using parametric processes (see Eqs.~\eqref{seq:phys.impl.hsr.Rsb} and \eqref{seq:phys.impl.hsr.Bsb}). The total system-reservoir dynamics can now be described by a Lindblad master equation 
$ (d \hat \rho _{\mathrm{sr}} / dt) =
\mathcal{L} 
_{ \mathrm{tot} }  
\hat{\rho} _{\mathrm{sr}} $, where the Lindbladian can be written as 
\begin{align}
\label{seq:phys.impl.Lsr}
& \mathcal{L} 
_{ \mathrm{tot} } 
\hat{\rho} _{\mathrm{sr}} 
= - i 
[\hat H _{\mathrm{sr}} 
 , 
\hat{ \rho} 
_{\mathrm{sr}} ] 
+ \kappa 
\mathcal{D} [ 
\hat z    ] 
\hat{ \rho}  
, \\
\label{seq:phys.impl.hsr}
& \hat H _{\mathrm{sr}} 
= \eta \gamma  
\hat x _{+} 
\hat y
+  \frac{ \sqrt{\gamma  
\kappa } }{ 2 } 
 [ 
\hat z ^{\dag}
\! \left( \hat x _{+} 
- i \eta \hat y \right ) 
+ \text{H.c.}   ] 
. 
\end{align}
In the limit $\kappa \gg \gamma  $, we can adiabatically eliminate the auxiliary mode to obtain the effective system master equation that is given by Eq.~\eqref{seq:qme.meas.ff}. Note that this approach also works for other kinds of auxiliary reservoirs, such as two-level systems, if we replace $\hat z $ and 
$ \hat z ^{\dag}$ with the spin lowering and raising operators, respectively. Alternatively, the dissipative dynamics in Eq.~\eqref{seq:qme.meas.ff} can also be realized using repetitions of interaction and resets~\cite{Blatt2011}.

As the Hamiltonian $\hat H _{\mathrm{sr}} $ in Eq.~\eqref{seq:phys.impl.hsr} is a sum of bilinear interaction terms, one can straightforwardly implement this part by engineering weak nonlinear elements between relevant modes, and then driving mode 
$c $ or $z$ with the appropriate drive tones that render the relevant sideband processes resonant. More specifically, we can decompose the Hamiltonian into red versus blue sideband processes, as 
\begin{align}
\label{seq:phys.impl.hsr.sb}
\hat H _{\mathrm{sr}} 
& = \hat H _{\mathrm{red}} 
+\hat H _{\mathrm{blue}} 
, \\
\label{seq:phys.impl.hsr.Rsb}
\hat H _{\mathrm{red}} 
& = 
\frac{\eta \gamma  }{ 2 }
\hat {m} _{+} 
\hat {c} ^\dag
+ \frac{ \sqrt{\gamma  
\kappa } }{ 2 \sqrt{2} } 
( \hat {m} _{+}  
- i \eta \hat c ) 
\hat z ^{\dag} 
+ \text{H.c.}  
, \\
\label{seq:phys.impl.hsr.Bsb}
\hat H _{\mathrm{blue}} 
& = 
\frac{\eta \gamma  }{ 2 }
\hat {m} _{+} ^\dag
\hat {c} ^\dag
+ \frac{ \sqrt{\gamma  
\kappa } }{ 2 \sqrt{2} } 
( \hat {m} _{+}  ^\dag
- i \eta \hat c ^\dag   ) 
\hat z ^{\dag}
+ \text{H.c.}  
, 
\end{align}
where we define 
$\hat {m} _{+} \equiv (\hat {a} + 
\hat {b}) / \sqrt{2} $.

It is important to note that the Hamiltonian in Eq.~\eqref{seq:phys.impl.hsr} dose not require any direct coupling between modes $a $ and $ b $, i.e.~the target modes that we want to entangle. This makes our protocol suitable for systems with constrained form of coupling, e.g.,~optomechanical systems where it is easier to engineer interaction between mechanical and optical modes or between optical modes, while a direct tunable mechanical coupling can be more challenging.

Finally, we note that the recovery step requires projection measurements of linear quadratures and local displacement operations. Both are standard operations in continuous variables systems and have also been demonstrated in a broad range of bosonic quantum systems including superconducting microwave resonators~\cite{Lehnert2008}, optical and acoustic cavities, and mechanical oscillators.

\subsection{Entanglement by purely dissipative QND dynamics}

\label{suppsec:ent.qnd.disp}

One can naturally ask which components of the dynamics in Eq.~\eqref{eq:qme.meas.ff} are essential for the entanglement generation. At first glance, it may be tempting to think that the Hamiltonian coupling between mode $a $ and register mode $ c$, which takes the form of a quantum-non-demolition coupling with $\hat H _{\mathrm{int} } 
= ( \eta \gamma /\sqrt{2} ) 
\hat x _{a} \hat y $, is essential for entanglement generation. Surprisingly, as shown in~\cite{Seif2022}, this is not the case: even purely dissipative processes can generate entanglement. For example, a single Lindblad dissipator as follows can already generate entanglement, with respect to a bipartition between $a$ versus ($b$, $c $) modes: 
\begin{align}
\label{eq:qme.dissp}
& \frac{d \hat \rho }{dt} 
= 2 \gamma  
\mathcal{D}\left[ 
\hat x _{+} 
- i \eta \hat y \right ] 
\hat{ \rho}  
\equiv 
\mathcal{L} _{\mathrm{disp} } 
\hat \rho  
. 
\end{align}

More interestingly, our analysis in the main text of Eq.~\eqref{eq:qme.meas.ff} can help elucidate the entanglement generation mechanism in Eq.~\eqref{eq:qme.dissp}. To see this, we first decompose Eq.~\eqref{eq:qme.dissp} into a sum of two measurement-and-feedforward processes in opposite directions, as  
\begin{align}
\label{eq:L0.gen.2way}
& \mathcal{L} _{\mathrm{disp} } 
= \gamma \mathcal{L} _{\hat x _{+} 
\to \eta \hat y } 
+ \gamma  \mathcal{L}_{ \hat y 
\to - \eta \hat x _{+} }  
    , \\
& \mathcal{L} _{\hat x _{+} 
\to \eta \hat y } 
\hat{ \rho} = 
- i  
[\eta \hat x _{+} 
\hat y , 
\hat{ \rho} ] 
+   
\mathcal{D}\left[ 
\hat x _{+} 
- i \eta \hat y \right ] 
\hat{ \rho}  
, \\
\label{eq:L0.gen.mff.local}
& \mathcal{L}_{ \hat y  
\to - \eta \hat x _{+} } 
\hat{ \rho} = 
 i  
[\eta \hat x _{+} 
\hat y , 
\hat{ \rho} ] 
+  
\mathcal{D}\left[ 
\hat x _{+} 
- i \eta \hat y \right ] 
\hat{ \rho}  
	. 
\end{align}
We now consider a quantum-trajectory realization of Eq.~\eqref{eq:L0.gen.2way}, where we first measure both the $\hat x _{+} $ and $\hat y $ quadratures, and based on the measurement results, apply conditional feedforward forces to 
$\hat y $ and $-\hat x _{+} $, respectively. During this procedure, only the nonlocal measurement of $\hat x _{+} $ can generate entanglement between the first and the other two modes. As such, the conditional state entanglement in Eq.~\eqref{eq:meas.cond.xp.ent} provides an upper bound on the final state entanglement. More intriguingly, one can show that the logarithmic negativity of the entanglement generated by Eq.~\eqref{eq:L0.gen.2way} exhibits the same asymptotic long-time behavior as the conditional state due to nonlocal measurement in Eq.~\eqref{eq:meas.cond.xp.ent}, i.e.~we have
\begin{align}
\label{seq:ent.disp}
t \to \infty: 
\, 
\mathcal{E} _{\mathcal{N}} 
( e ^{ t \mathcal{L} _{\mathrm{disp} } }
\hat \rho _{\mathrm{i}} ) 
\sim \frac{1 }{ 2 } 
\log \left ( 
\gamma t \right)
+ o \left ( 
1 \right)
. 
\end{align}

\end{document}